\documentclass[10pt,journal]{IEEEtran}
\usepackage[utf8]{inputenc}
\usepackage{graphicx}    
\pdfoutput=1
\usepackage{amsmath}
\usepackage{chngcntr}
\usepackage{array}
\usepackage{epsf}
\usepackage{multirow}
\usepackage{rotating}
\usepackage{caption}
\usepackage{epsfig}
\usepackage{subcaption}
\usepackage{epstopdf}
\usepackage{algorithm}
\usepackage{algorithmic}
\usepackage{xcolor}
\usepackage{cite}
\usepackage{float}

\RequirePackage[normalem]{ulem} 
\RequirePackage{color}\definecolor{RED}{rgb}{1,0,0}\definecolor{BLUE}{rgb}{0,0,1} 

\title{Cooperation-based Routing in Cognitive Radio Networks}
\author{\large Arsany Guirguis$^\ddagger$, Mohammed Karmoose$^\dagger$,
  Karim Habak$^\star$, Mustafa El-Nainay$^\ddagger$ and Moustafa Youssef$^{\star \star}$\\ [.1in]
\normalsize

\begin{tabular}{c}\
$^\ddagger$Computer and Systems Engineering, Alexandria University, Alexandria,
Egypt\\
$^\dagger$Department of Electrical Engineering, Alexandria University,
Alexandria, Egypt. \\
$^\star$School of Computer Science, College of Computing, Georgia Tech.,
Atlanta, Georgia, USA\\
$^{\star \star}$Egypt-Japan Univ. of Sc. and Tech. (E-JUST), Alexandria, Egypt\\
Email: arsany.guirguis@alexu.edu.eg, m\_h\_karmoose@alexu.edu.eg,
karim.habak@cc.gatech.edu,\\ ymustafa@alexu.edu.eg, moustafa.youssef@ejust.edu.eg \end{tabular} } 

\def \sys {\textit{Undercover}}
\begin{document}
\maketitle

\begin{abstract}
Primary user activity is a major bottleneck for existing routing protocols in
cognitive radio networks. Typical routing protocols avoid areas that are highly
congested with primary users, leaving only a small fragment of available links for
secondary route construction. In addition, wireless links are prone to channel
impairments such as multipath fading; which renders the quality of the available
links highly fluctuating. In this paper, we investigate using cooperative communication 
mechanisms to reveal new routing opportunities, enhance route qualities,
 and enable true coexistence of  primary and secondary networks.
 As a result, we propose Undercover: a cooperative routing protocol that
utilizes the available location information to assist in the routing process. Specifically, our protocol revisits
a fundamental assumption taken by the state of the art routing protocols designed for cognitive radio
networks. Using \sys{}, secondary users can transmit in the regions of primary users 
activity through utilizing cooperative communication techniques to null out 
transmission at primary receivers via beamforming. In addition, 
the secondary links qualities are enhanced using cooperative diversity. To account for the
excessive levels of interference typically incurred due to cooperative
transmissions, we allow our protocol to be interference-aware. Thus, cooperative
transmissions are penalized in accordance to the amount of negatively affected
secondary flows. We evaluate the performance of our proposed protocol via NS2
simulations which show that our protocol can enhance the network goodput by a
ratio reaches up to 250\% compared to other popular cognitive routing protocols
 with minimal added overhead.
\end{abstract}

\section{Introduction}
Cognitive Radio Networks (CRNs) are imminent to pervade into all fields of
wireless communications. We backup this assertion with three observations.
First, with the inherent inefficiency of the static spectrum licensing
policies~\cite{force2002report} and proliferation of spectrum accessing mobile
and IoT devices~\cite{IoTRef}, we are quickly heading towards a wireless
spectrum crisis~\cite{crisis}. Second, the industrial market is widely shifting focus towards CRNs: software defined radio technologies are expected to dominate the market in the
coming years, with a \$27.29 Billions net worth of a market size by the year
2020~\cite{SDRMarket}. Finally, the FCC announced new regulations allowing for
wireless spectrum reuse by unlicensed users~\cite{FCC-ruling}. These
regulations allow the unlicensed secondary users (SUs) to access the spectrum
as long as they do not interfere with the licensed primary users (PUs).

One of the most challenging functionalities in cognitive radio networks is
routing. Routing protocols for CRNs have attracted attention of a large number of researchers \cite{crn_metric_survey}.
In these networks, the routing protocol's logic aims to determine (1) the
route to the destination and (2) the channels to be used along this route. More
importantly, it ensures that the used routes do not interfere with the licensed
PUs. Therefore, the research community has explored various forms of
routing in cognitive radio networks
\cite{youssef2014routing,pefkianakis2008samer,chowdhury2009search,habak2013location}.
These explorations led to the development of various routing protocols that suit these goals.

One common assumption in existing routing protocols, however, causes a critical limitation of the routing process and therefore hinders CRNs to perform suitably for real-world applications. Specifically, the current state-of-art
cognitive radio routing protocols assume that {\it SUs inside the
interference range of a certain PU can not utilize its licensed channel
during its active periods}. This assumption is too constraining in comparison
to the FCC's regulations, which {\bf only} prevent SUs from
interfering with the licensed PUs~\cite{FCC-ruling}. Such
constraining assumption leads to (1) wasting possible communication
opportunities, (2) relying on relatively suboptimal routes due to PUs activities, and (3) being unable to construct routes in scenarios where
PUs are highly dense and/or relatively active. The question then arises: ``Can SUs {\em utilize} licensed channels that are occupied with active PUs, {\em without} interfering with them?''

A partial answer to this fundamental question is found in the literature of physical layer communications. Cooperative communications \cite{goldsmith2005wireless,tse2005fundamentals} have been extensively studied as a mean to enhance the reliability of communication links.
By allowing transmitters/receivers to cooperatively encode/decode transmitted/received data, communication links can be rendered less vulnerable to negative communication environments such as poor communication channels and excessive interference levels.
One particular technique is Cooperative Beamforming \cite{van1988beamforming}, where precoding is employed cooperatively by a set of transmitters to null out transmission at particular directions of interest, while simultaneously combating poor channel conditions to increase the reliability of transmission links.
Cooperative beamforming can therefore be well-fitted for overcoming the aforementioned assumption: by allowing SUs to cooperatively null-out transmission at the directions of active PUs, while maintaining secondary transmission at the required level. Thus, physical layer communication mechanisms provide the means for truly undercover communication that enables SUs to communicate without interfering with PUs. Although being well-established in the literature, cooperative beamforming has been solely considered in Cognitive Radio setups which involve {\bf single-hop} transmission links\cite{chen2014cooperative,tao2012overview,karmoose2013stability,yi2012cooperative}. First steps have been taken towards developing multi-hop-based cooperative beamforming schemes. However, DZP
\cite{karmoose2013dead} is the first to consider using cooperative beamforming
in the routing process. But, this protocol only uses such mechanism
for route maintenance as a reaction to the PU activity not as a routing protocol.

In this paper, we take a second step towards answering that question. We investigate the possibility of employing cooperative beamforming
techniques in CRNs in a multi-hop context to enables concurrent primary and secondary transmissions in the same geographical area.
We develop a cross-layer based routing protocol, {\it \sys{}}, that utilizes the available location information to route
data towards their destination. Our proposed protocol enables SUs to
construct cooperative groups that employ cooperative beamforming to either enhance the attained secondary throughput or
to null out reception at nearby PUs, thus allowing for simultaneous use of the
spectrum by both PUs and SUs. Considering the elevated levels
of interference commonly exhibited by cooperative transmissions, we allow our
protocol to be interference-aware by penalizing the routing situations which
may incur excessive interference on on-going secondary flows.  Our proposed
protocol is considered as a local one: each node is provided with the necessary
information about the network topology and wireless channel condition. This
allows for a fully distributed implementation of the protocol.

We evaluate the performance of \sys{} via NS2 simulations~\cite{NS2}  under various network
conditions. We compare \sys{}'s achieved performance against other cognitive
radio routing protocols. Our experiments show that \sys{} achieves
up to 250\% increase in goodput compared to other protocols. In addition, the done experiments show that the path construction overhead introduced by \sys{} enables
relying on stable paths that avoid the rerouting overhead triggered by
PUs activity.

Overall, the contributions of this paper are as follows:
\begin{enumerate}


\item Designing \sys{}, a cooperative based routing protocol that operates on the new opportunities and uses location information to aid the routing process. It utilizes the idea of beamforming to enable primary and secondary transmission to co-exist, by allowing SU to null transmission in the directions of existing PUs. It also uses cooperation with neighboring nodes to increase secondary throughput.


\item Formulating the tradeoffs introduced by the need to enhance the path communication throughput while minimizing the inter-path interference. For this, we analyze the average achievable capacity using a single node and a cooperative group in the routing process. We also develop a routing metric that is based on the achievable capacity of single, as well as multiple transmitters employing beamforming for a single-hop setup

\item Developing heuristics to shrink the search space of choosing nodes to participate in the coopeartive group. For this, we propose a group selection algorithm that provides candidate cooperative groups of SUs, which operates in small time that allows for practical usage.

\item Evaluating \sys{} performance using NS2 simulations against other popular routing protocols that are designed for cognitive radio networks.

\end{enumerate}

The rest of the paper is organized as follows. Section \ref{sec::related_work}
presents some background material in addition to our related work. We then
describe our system model in Section \ref{sec:sysModel}.  We present our
routing objectives and propose our routing metric in Section
\ref{sec:routeMetric}. We then describe the whole routing protocol in Section
\ref{sec::routing_protocol}. Section \ref{sec::eval} evaluates our proposed
routing protocol and Section \ref{sec::conc} concludes the paper and provides
directions for future work.

\section{Background and Related Work}
\label{sec::related_work}
The wireless spectrum is a broadcast medium where different radio waves can
constructively or destructively collide and affect one another. The research
community has thoroughly investigated utilizing these effects to alter the
signal transmission characteristics by allowing multiple users to transmit
carefully selected signals simultaneously and collaboratively
\cite{tao2012overview,goldsmith2005wireless,tse2005fundamentals}. These
investigations have led to enhancing communication throughput
\cite{lakshmanan2009diversity}, enabling simultaneous non-interfering
transmissions \cite{collier1991transmission}, and reshaping the transmission
beam \cite{van1988beamforming}.

There have been some attempts to exploit cooperative communication in the context of
{\bf conventional wireless networks}. These attempts were led by Khandani et al.  who
analyzed the
energy saving benefits introduced by employing cooperative diversity
\cite{khandani2005cooperative}. However, despite the sound theoretical
framework, their proposed algorithms are not suitable for real-time
communication. With similar energy efficiency goals, other researchers proposed
 more real-time suitable mechanisms in static environments with wireless
shadowing\cite{khandani2007cooperative,li2006energy} as well as in multi-path
environments\cite{ibrahim2008distributed,jakllari2007cross}. Other researchers
focused on enhancing the network throughput in case of having multiple
concurrent flows
\cite{lakshmanan2009diversity,lakshmanan2012proteus,sharma2010cooperative}.

The significant performance gains introduced by cooperative diversity have
motivated some researchers to employ its techniques in the context of {\bf cognitive
radio networks} \cite{zhang2009cooperative,ding2010distributed}.  For instance,
Ding et al. use cooperative diversity to transmit data with higher capacity to
maximize throughput \cite{ding2010distributed}. In addition, Sheu et al.
proposed a cooperative routing protocol in \cite{sheu2012cooperative} that enhances the end-to-end
throughput. Unfortunately, non of these approaches (1) offer new communication
opportunities, (2) mitigate the effect of having active primary users on the
secondary network, (3) address the inter-path interference problem which
increases as a result of cooperative transmission.

One of the intensively studied cooperative communications techniques is
cooperative beamforming, which relies on sending  precoded versions of the same
data to reshape the signal beam  producing transmission nulls at certain
spatial directions \cite{van1988beamforming}. A direct consequence of employing
cooperative beamforming is to allow for spatial multiplexing of concurrent
transmissions of multiple nodes \cite{tse2005fundamentals}. Fortunately,
cooperative beamforming provides the means for hiding secondary user
communication from primary users and avoiding interfering with primary user
communication. This opportunity was considered by a small number of
attempts like \cite{yi2012cooperative} which utilizes beamforming by developing MAC layer protocols that maximize
the received signal-to-noise ratio (SINR) among SUs with different power
constraints and the QoS requirement of PUs. However, this protocol deploys beamforming among {\bf one} relay node only between the source and the destination.

Our previous work \cite{karmoose2013stability} considered using beamforming in the routing layer. However, we only
proposed a route maintenance mechanism to alleviate the need for route
re-establishment upon the detection of a PU which limits the usefulness of
beamforming. In this paper, we consider cooperative beamforming a foundation
for building successful cognitive radio routing protocols. We understand the
true potential of employing cooperative beamforming, mitigate its introduced
interference effects, and provide practical mechanisms for efficient
cooperative group construction.


\begin{figure*}
\begin{tabular}{ccc}

 \begin{minipage}{0.3\textwidth}
 \centering
\begin{figure}[H]
\centering
 	\begin{subfigure}[t]{1\textwidth}
	\centering
    \includegraphics[width=1\linewidth]{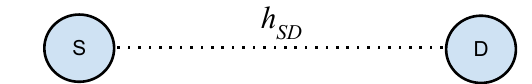}
	\caption{Node-to-node case.}
	\label{fig::n2ncase}
	\end{subfigure}
	\begin{subfigure}[t]{1\textwidth}
	\centering
    \includegraphics[width=1\linewidth]{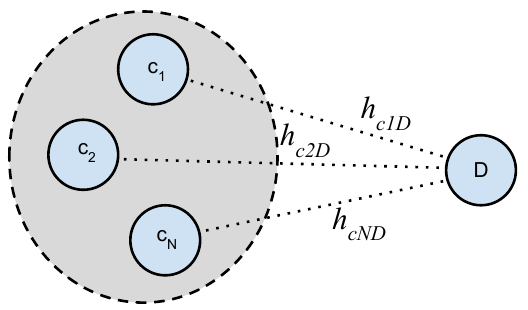}
	\caption{Multi-node-to-node case.}
	\label{fig::g2ncase}
	\end{subfigure}
\caption{Existing options to transmit data. This is used to clarify throughput calculations.}
\label{fig::capacity_calculations}
\end{figure}
 \end{minipage}
 &
 \begin{minipage}{0.3\textwidth}
 \centering
 \includegraphics[width=2.2 in]{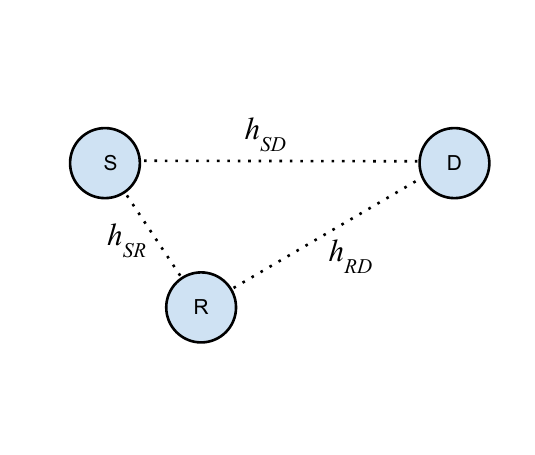}
 \captionsetup{justification=centering}
 \caption{A scenario showing the difference between sending as a single node or as a group of two nodes.}
 \label{fig::relay}
 \end{minipage}
 &
 \begin{minipage}{0.3\textwidth}
 \centering
 \includegraphics[width=2.4 in]{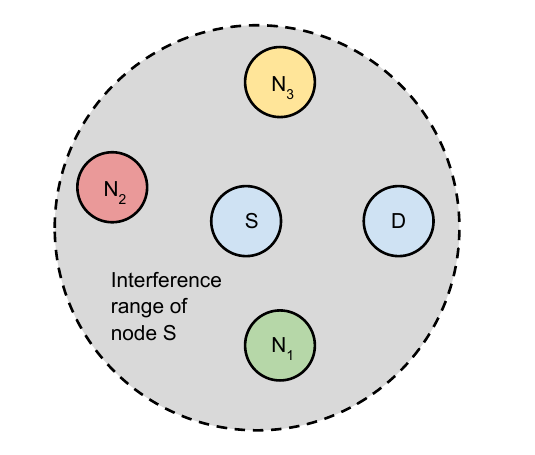}
 \caption{Illustrative example of interference effect.}
 \label{fig::interference_example}
 \end{minipage}

 \end{tabular}
 \end{figure*}
\section{System Model} \label{sec:sysModel}
In this section, we present our system assumptions and then provide a brief overview on the proposed routing protocol.

\subsection{System Assumptions}
Throughout this work, we assume the presence of a CRN which consists of a set of
stationary PUs that are licensed to use the spectrum according to
their data delivery requirements. This is common in many CRN scenarios such as
TV white space-based CRNs. We further assume that each SU knows its own location and the location of its direct neighbors. A node can estimate its location using any of the current localization systems, including GPS \cite{GPS99} or cellular mobile-based systems \cite{GSM_loc4}. Moreover, a sender can obtain the location information of the ultimate destination via out of band services that map node addresses to locations, or have it disseminated through the network.
Assuming knowing only the location of the destination is a typical and valid assumption in many applications including military and sensor networks where reporting nodes know the locations of the sink nodes.
 We assume also that SUs are able to sense
and detect the PUs activity \cite{punch}. A set of stationary SUs are allowed to
use the spectrum in a manner that does not jeopardize the integrity of the PU
transmissions, with maximum transmission power of $P_T$ for each SU. Assuming
that the primary network adopts an overlay transmission policy, a SU is not allowed to transmit except in two cases: 1) a PU is not currently
active, in which case a SU is permitted to occupy the spectrum for a period
which terminates with the re-occupation of a PU to the spectrum, or 2)
concurrently with the PU only if a SU is able to employ cooperative
beamforming. It is vital to mention here that interference must be avoided at
the receiving end of the primary link. Therefore, we assume throughout this
work that a PU signifies a primary receiver, and that concurrent transmission
is only allowed if interference is avoided at the receiver side. We assume also
a slow fading multi-path wireless channel, in which a channel coefficient is
constant over a period of $T_c$. A short $T_c$ would require frequent
estimation of the channel coefficients, while a long $T_c$ alleviates this
requirement.

Primary receiver detection can be done via overhearing and decoding reply
messages sent by the receiving nodes, such as ARQ or CTS packets. Assume that
PUs are separately identified by each of the SU nodes in the network. Such
an assumption can be validated by observing that decoding ARQ or CTS packets
sent by PUs. This gives the SU identifying information of the PU, such as the MAC
address of its equipped NIC card. In order to perform accurate beamforming,
reliable estimates of the channel coefficients between the transmitting SUs and
the PU should be acquired, and this can be achieved by employing channel
estimation techniques on the preambles of those packets\cite{de1997maximum}.
SUs are assumed to communicate control packets over a dedicated Common Control
Channel (CCC) such as the 2.45GHz ISM band \cite{khop}.

\subsection{Routing Protocol Overview}
We assume that our protocol will be deployed in a cognitive radio network that employs an overlay transmission policy. In such network, PUs and SUs cannot transmit signals simultaneously except via cooperative beamforming. When some node has a data packet to send, it broadcasts a route request (RREQ) packet to all of its neighbors and waits for replies. First, each neighbor is checked for its ability to host the data packet to the target destination. A neighbor is able to be a part of the route if its distance to the final destination is less than the distance from the source to the destination. In this case, such neighbor is considered as a potential relay node for this transmission. Each potential relay considers all of its neighbors (other than the source node) as possible next hops. For each possible next hop, the relay tries to construct different groups of different sizes with its neighbors (excluding the potential next hop). For each potential constructed group, its routing metric is calculated and saved. After trying all of the possible groups, the relay node finds the best group routing metric and sends it back as a route reply (RREP) packet to the requester node.

A source originally generating a RREQ waits for a timeout period during which it collects received RREP from neighboring nodes. Upon its termination, a node chooses the link with the highest link metric as the next hop, to which it sends an ACK packet. As the next hop receives the ACK packet, it knows that it has been decided for the forwarding of the source packet from the originating source. It then sends an Ack Reply (AREP) packet to the source. Finally, it receives the data packet and becomes responsible for disseminating it to the constructed group so that all participating nodes in this group can send it to the already chosen next hop. Such node, in this case, is called the group coordinator. When the next hop receives the data packet, it repeats again the same algorithm to find the next hop on the route. This procedure is repeated until the data packet reaches its final destination. 
\begin{table}[h]
\centering
\caption {Mathematical Notations}
\label{mathnot}
\begin{tabular}{|p{1cm} |p{6.5cm}|}
\hline
Symbol                          & Description    \\ \hline 

$P_T$                   &  Maximum transmission power for SUs           \\ \hline
$P_c$                   &  Received power by some node due to the transmission of a cooperative group           \\ \hline
$P_{int}$             &  Interference limit of SUs           \\ \hline
$P_{FSP}$           &  The level of interference caused by a transmitting node           \\ \hline
$h_{sd}$             &  Channel coefficient between two nodes $s$ and $d$           \\ \hline
$w_{ij}$              &  Beamforming weights between two nodes $i$ and $j$           \\ \hline
$\sigma_N^2$                   & Variance of Gaussian noise that affects each node in the network          \\ \hline
$B$                   & Available bandwidth for some particular link           \\ \hline
$C_{sd}$                   &  Achievable capacity between nodes $s$ and $d$           \\ \hline
$C_{cd}$                   &  Achievable capacity between cooperative group $c$ and node $d$           \\ \hline
$\hat{C}_{ij}^{coop}$                  &  Achievable capacity during cooperative transmission from $i$ to $j$           \\ \hline
$C_{ij}^{wor}$                   &  The worst capacity achieved to deliver the packet from the group coordinator to any of the neighbors participating in some cooperative transmission          \\ \hline
$\hat{C}_{ij}$                   &  The effective capacity achieved through sending as a group           \\ \hline
$N_f$                   &  Number of on-going flows that are in the interference range of the cooperative group           \\ \hline
$N_n$                   &  Number of flow-carrying direct neighbors of all the participating nodes in the transmitting group           \\ \hline
$N_{min}$                   &  The value of $N_n$ for the minimum allowable group size           \\ \hline
$d_r$                   &  Radius of the circle describing the effective interference range of some group            \\ \hline
$A$                   & Area covered by nodes participating in some group  \\ \hline
$D_f$                   &  Density of the flows surrounding some group           \\ \hline
$D_n$                   &  Node density around some node based on the two-hop neighbor information           \\ \hline
$F_n$                   &  A rough estimate of the number of flows per node at nodes surrounding some group           \\ \hline

\end{tabular}
\end{table}

\section{Mathematical Model} \label{sec:routeMetric}
In this section, we give the proposed mathematical formulation of our proposed routing metric. First, we give a brief overview about the metrics that affect the routing decisions and the incentives behind them. Then, we give the details of the mathematical model for each of them. Finally, the complete routing protocol metric is given.
\subsection{Overview}
The proposed routing protocol aims at maximizing the achievable throughput across the network. Links along the route are chosen based on the maximum achievable capacity. However, in order to account for throughput calculations along the links, a node should be aware of the sources inflicting interference and therefore reducing the achievable capacity when transmitting to this node. A precise calculation of the interference level due to \textbf{all} transmitting sources is a difficult task, since it is now possible for two or more nodes that are not in the interference range of the receiving node to inflict interference if they engage in a cooperative transmission phase. In other words, cooperating groups in the network can cause considerable interference levels at relatively distanced nodes that are unaware of this source of interference.

The incentive in adding the interference terms to the routing metric is to penalize the use of cooperative groups in a manner that is proportional to the amount of expected interference incurred at nearby concurrent flows. The proposed metric is a composite of two components: the first gives an estimate of the achievable throughput across the links, and the second captures the interference effect of using a cooperative group across the network. At this point, we make the assumption that the source node is \textbf{not} responsible to deliver the packet to all the cooperating nodes in the next hop. It is only required to deliver the data packet to the \textit{group coordinator} which is responsible for data dissemination among the collaborating nodes. This should be taken into account in the calculation of the cost. Table \ref{mathnot} summarizes all mathematical notations that are used throughout this paper.

\subsection{Capacity Calculations}
\label{capacity_calculations}
In this section, we present the mathematical formulation of the achievable throughput across a link. As shown in Figure \ref{fig::capacity_calculations}, there are two possible links that can be utilized along a given route. Those are: 1) a node-to-node link and 2) a multi-node-to-node link. We calculate the maximum achievable throughput through utilizing each of these two link types. The following discussion is primarily based on basic wireless communication concepts that can be found in \cite{goldsmith2005wireless,tse2005fundamentals}. In all cases, we assume that each node is affected by thermal additive white Gaussian noise of variance $\sigma_N^2$.

\paragraph{Node-to-node link}\label{n2n} Assume that nodes $S$ and $D$ are the transmitter-receiver pair of a simple link and that the transmitted signal is affected by slow multi-path fading. Let $h_{sd}$ represents the multi-path channel coefficient between $s$ and $d$. The achievable capacity between nodes $S$ and $D$ can be calculated as

\begin{equation}
\begin{split}
 \label{capacity_simplex}
 C_{sd} &= B \log (1 + \text{SINR}_{sd}) \\
 \text{SINR}_{sd} &= \dfrac{P_T \| h_{sd} \|^2}{\sigma_N^2}
 \end{split}
\end{equation} 

\noindent where $B$ is the available bandwidth for this particular link.

\paragraph{Multi-node-to-node link} Assume that nodes $c_1$ to $c_N$ cooperatively send data to node $d$ in the presence of a set of primary receivers $P = \{P_1,P_2, \hdots , P_M\}$. It is important to note here that $N$ should be strictly larger than $M$; the number of group members should be greater than the number of surrounding PUs. The cooperative group then applies the appropriate beamforming weights $\bar{w}_{cP} = [ w_{c_{1 P}} \hdots  w_{c_{N P}} ]$ to null transmission at the primary receivers while maximizing the achievable throughput at node $d$. The achievable capacity in this case becomes

\begin{equation}
\begin{split}
\label{capacity_cooperative_group}
 C_{cd} &= B \log( 1 + \text{SINR}_{cd}) \\
 \text{SINR}_{cd} &= \dfrac{P_{c}}{\sigma_N^2}, \:  P_{c}  = P_T \| \bar{w}_{cP}^H \bar{h}_{cd} \| ^2
\end{split} 
\end{equation} 

\noindent where $x^H$ denotes the complex hermitian of a vector $x$, $\bar{h}_{cd} = [ h_{c_1 d}^H \hdots  h_{c_N d}^H ]^H$ is the channel coefficients vector between the nodes of the cooperative group $c_1$ to $c_N$ and $d$, and $P_{c}$ is the received power by node $d$ due to the transmission of the cooperative group. Note that for appropriate transmission nulling, the following constraint must be satisfied
\begin{equation}
 \bar{w}_{cP} \in \text{Null}({\bf H}_P)
 \end{equation} 
 \noindent where $\text{Null}(X)$ denotes the null space of matrix $X$ and ${\bf H}_P$ is a matrix whose rows are $\bar{h}_{c P_j} = [h_{c_1 P_j} \hdots  h_{c_N P_j}]$ for $j=1,\hdots,M$ where $h_{c_i P_j}$ is the channel coefficient between node $i$ and primary receiver $j$.

\paragraph{Effective Capacity}

Our proposed metric consists of the maximum achievable capacity between node i and j based on knowledge of channel coefficients. Consider the scenario in Figure \ref{fig::relay}. Node S wants to deliver some packet to node D through R which does not know the packet yet. Assume packet length $L$. Node S sends the packet to node R with capacity $C_{SR}$. The time needed to deliver the packet is $T_{SR} = L/C_{SR}$. After receiving, cooperation will take place, and the achievable capacity will be $\hat{C}_{SD}$ which is greater than $C_{SR}$. The time needed to deliver the packet will be $T_{SD} = L/\hat{C}_{SD}$. The effective capacity in this case will be $\hat{C} = L/(T_{SD}+T_{SR}) = \hat{C}_{SD} C_{SR}/(\hat{C}_{SD} + C_{SR})$. For a relatively large discrepancy between $\hat{C}_{SD}$ and $C_{SR}$, the effective capacity can be written as $\hat{C} \simeq \min (\hat{C}_{SD},C_{SR})$. The preceding analysis can be readily extended to any multi-node-to-node transmissions from node i to j, for which the effective capacity is 
\begin{equation}
\hat{C}_{ij} \simeq \min (\hat{C}_{ij}^{coop},C_{ij}^{wor})
\end{equation}
where $\hat{C}_{ij}^{coop}$ is the achievable capacity during cooperative transmission from i to j, and $C_{ij}^{wor}$ is the minimum capacity achieved to deliver the packet to any of the neighbors participating in this cooperative transmission.

\subsubsection*{Discussion}
As witnessed from the previous analysis, the process of disseminating the data packet to the group members prior to the transmission phase can be an enervating factor to the achievable effective capacity. However, two notes should be pinpointed: 1) despite the aforementioned obstacle, a cooperative group of such structure can provide better throughput than conventional point-to-point links. Consider the following scenario: $h_{SD} = 2$, $h_{SR} = 2\sqrt{2}$ and $h_{RD} = 2$. Assuming unity transmission power and noise variance, the achievable capacity through the point-to-point links S-D, S-R and R-D respectively are $\log(5)$, $\log(9)$ and $\log(5)$. We can see that direct routes from S to D are all hindered by the bottleneck throughput of $\log(5)$. However, if nodes R and S are to cooperatively send the data to node D, then a maximum throughput across this multipoint-to-point link is $\log(9)$. In this example, despite having to report the packet to the cooperating node through a capacity-limited channel, the 
overall performance of the two-hop communication is superior to any of the point-to-point links available. 2) There are situations in which multiple nodes among the cooperative group are informed by the packet-to-be-delivered by overhearing the source transmission. This situation is commonly favorable because it alleviates the need for intra-communications among the cooperating nodes and therefore achievable capacity limits can be increased. Consider the previously mentioned scenario: if node R was to overhear the packet, then the throughput attained across the multipoint-to-point link, where S and R collaboratively send to D, is $\log(9)$ in contrast to $\log(5)$ attained over conventional routes.

\subsection{Interference Calculations}
In this section, we study the interference results from sending data through a cooperative group. We analyze two types of interference which are interference on the group due to its neighbors and the interference due to the group on the neighboring nodes.


\subsubsection{Interference due to neighbors}
As was previously mentioned, transmitting nodes in the same interference range are able to use the spectrum in a contending fashion, which is generally resolved by multiple access techniques such as CSMA/CA. The number of direct neighbors contending for the spectrum directly affects the average achieved capacity for a transmitting node. This is demonstrated in Figure \ref{fig::interference_example} where node S wants to communicate with node D in the presence of three flow-carrying nodes in its interference range. Let $C_{SD}$ be the achievable capacity along the direct link between the two nodes in the presence of no interfering sources. Assuming the fair distribution of the time allocation for the medium among the 4 nodes, then we can state that the average achievable capacity between S and D in the presence of the interfering nodes is $C_{SD}/4$. The same argument holds in case a cooperative group is utilized instead of a single transmitting node. This concludes that the number of flow-carrying nodes in the 
interference range of the source node/group is inversely proportional to the average achievable capacity over the transmission link.

\subsubsection{Auto-interference due to cooperative groups}
A cooperative group inflicts a relatively higher level of interference on nearby nodes. Moreover, the effective interference range of a cooperative group is larger than that of a single node, which imposes extra difficulties in the design of an efficient interference-aware routing protocol. The reason for this is that conventional contention-handling protocols are oblivious to these out-of-range transmissions and thus have no control over them.

In order to alleviate this shortcoming, the proposed metric allows each transmitting node/cooperative group to keep track of the interference inflicted by its own transmission. Depending on the interference level, the use of this node/cooperative group is penalized which affects the route decision. A node should keep track of the on-going flows in its neighborhood; each node reports its witnessed flow IDs to its neighbors. Based on the received information from all neighbors along with neighbors known locations, a node can build a statistical model of the flow density in the surrounding area. We can estimate the inclusive area of the nodes by a general polygon whose vertices are the neighboring nodes. Assuming the obtained estimate area is $A$, the flow density is thus
\begin{equation}
 D_f = \dfrac{\text{Total number of distinguished flows}}{A}
\end{equation} 

\begin{table*}[ht]
\centering
\large
   \begin{tabular}{|ccc|c|c|c|c|c|c|c|c|c|}
   \hline
  \multicolumn{3}{|c|}{Group Size} & 2 & 3 & 4 & 5 & 6 & 7 & 8 & 9 & 10 \\ 
  \hline 
  \multicolumn{1}{|l}{\multirow{3}{0.7 in}{{ Accepted \% Inc. in $N_n$}}} & \multicolumn{1}{l|}{} & 1 PU & 85.6 & 148.8 & 197 & 236 & 268.9 & 298.2 & 321.7 & 342.2 & 362.1 \\
  \cline{3-12} 
  \multicolumn{1}{|l}{} & \multicolumn{1}{l|}{} & 2 PU &  & 36.3 & 75.5 & 111.5 & 145.4 & 174.1 & 199.1 & 220.7 & 241.8 \\
  \cline{3-12} 
  \multicolumn{1}{|l}{}& \multicolumn{1}{l|}{} & 3 PU &  &  & 20.3 &  44.2 &  71.3 &  97.4 & 122.1 & 145.5 & 167.4 \\
  \hline
  \end{tabular} 
  \caption{Acceptable percentage increase in $N_n$ for different group sizes relative to minimum allowable group size. The presented values are based on Rayleigh channel model with unit variance.}
  \label{table_threshold}
\end{table*}

In order to determine an estimate for the number of on-going flows that are affected by the transmission of a cooperative group, the effective interference range of the group should be also estimated. For simplicity, we assume that the far transmissions are affected by Free-Space-Path (FSP) loss. The level of interference caused by a transmitting node with power $P_T$ at node situated at distance $d$ can be approximated by

\begin{equation}
  P_{FSP} = P_T \dfrac{c}{d^2}
\end{equation} 

\noindent where $c$ is the free space path loss and its value depends on the used frequency. Given an interference limit of secondary nodes $P_{int}$, a cooperative group consisting of $N$ nodes can then calculate its effective interference range which is defined as: \textit{the geographical area in which a secondary node is -in the worst case- affected by an interference power greater than $P_{int}$, given the assumed FSP model}. Assuming the worst case scenario (perfectly coherent addition of transmitted signals at any given point), the received power at any point $(x,y,z)$ in space due to cooperative group transmission is given by

\begin{equation}
\label{interference_range}
 P_r(x,y,z) = P_T \sum_{i=1}^N \dfrac{\|w_i\|^2}{d_i^2}
\end{equation} 
\noindent where $d_i$ is the distance between node $i$ in the group and the point $(x,y,z)$. For ease of calculations, we approximate the effective transmission range of the cooperative group as if a single node situated in the center of the group, and transmitting with a power equal to $P_T \sum_{i=1}^N \|w_i\|^2$. Accordingly, the effective interference range is simply a circle of radius \begin{equation}
\label{interference_radius}
 d_r = \sqrt{\dfrac{c P_T \sum_{i=1}^N \|w_i\|^2}{P_{int}}}                                                                                                                                                                                                                                                                                                                                                                                                                                                                  \end{equation} 

The expected number of affected on-going flows $N_f$ can now be calculated by $N_f = D_f \times \pi d_r^2$.
  
\subsection{Proposed Metric}

Based on the above, the link metric between nodes $i$ and $j$ can be formulated as follows:

\begin{equation}
\label{metric}
LC_{ij} = \dfrac{\hat{C}_{ij}}{N_n + \beta (N_f - N_n)}
\end{equation} 

\noindent where $C_{ij}$ is the maximum achievable capacity between node $i$ and $j$ among all possibilities of transmission (either a single transmitting source or cooperative groups), $N_n$ is the number of flow-carrying direct neighbors of all the participating nodes in the transmitting group, $N_f$ is the number of on-going flows that are in the interference range of the cooperative group (this is equal to 0 for a single transmitter), and $\beta$ is a design parameter to alter the altruistic/egoistic behavior of the cooperative group. It should be noted that $N_f$ is inherently inclusive to the flows witnessed by the nodes in the cooperative group. That is the reason for the subtraction in (\ref{metric}).
\section{Implementation Details}
\label{sec::routing_protocol}
This section gives some practical and implementation details of the whole process. First, we present some practical issues that we consider for our routing protocol. Then, we give the details of the information exchanged among the nodes to serve the routing process. Finally, we present the whole flowchart of our algorithm along with an example that highlights how \sys{} works.
\subsection{Practical Considerations}
 \label{sec::link_formation}
 Our proposed protocol chooses the best possible link formation (with maximum link metric) among all possibilities of cooperative group constructions. Specifically, a source node sends a route request to its neighbors. Each of the neighbors calculates the link metric for all possibilities of transmission to determine which group is the best, and then sends a reply message to the source node for selection. The details of the protocol are described in Section \ref{sec::routing_protocol}. However, we discuss here the details of the link calculation algorithm.
 
 A node searches for the best link construction by calculating the link metric for all possibilities of cooperative transmissions. These possibilities are all based on the inclusion of direct neighbors of the potential relay node in a possible cooperative group. Assume the relay node has $N$ direct neighbors, then calculating the metric for all possible combinations of groups is $\mathcal{O}(2^N)$. However, if we assume that the node calculating the link metric is in the interference range of $M$ active PUs, then any allowed transmission should include at least $M+1$ cooperative nodes that are also in the interference range of these PUs. In other words, the number of nodes participating in the cooperative group should be strictly higher than the number of surrounding PUs ($N>M$). In this case, the complexity of the process of metric calculations will be reduced. Nonetheless, the current realization of the metric calculations is of a relatively high complexity. In the next discussion, we study the statistical characteristics of the proposed metric, based on which we reduce the complexity of the link calculation algorithm via less-probable candidates elimination.
 
  \begin{figure*}
\begin{tabular}{ccc}

 \begin{minipage}{0.5\textwidth}
 \centering
 \includegraphics[height = 2.2 in]{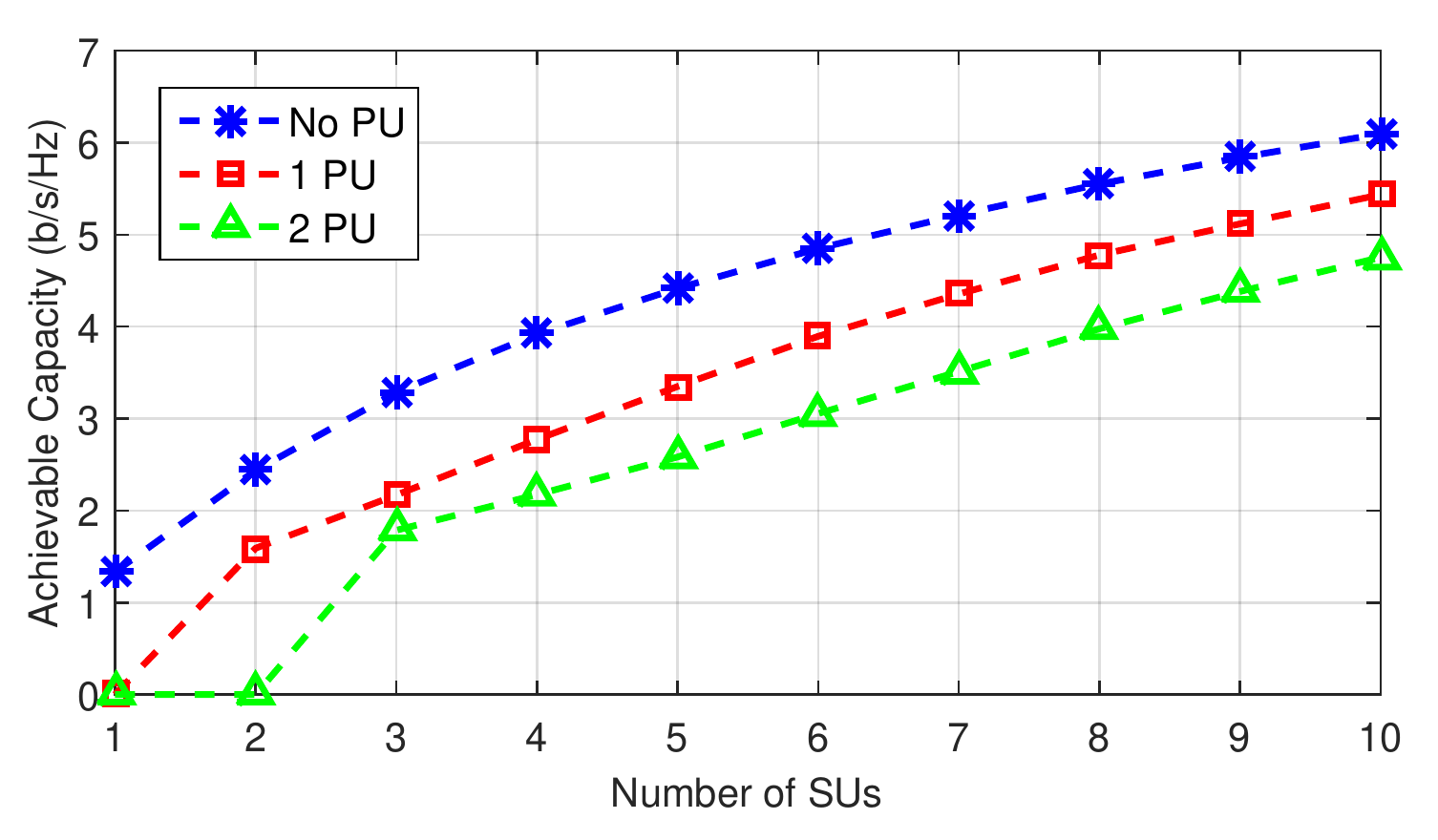}
 \caption{The effect of changing number of SUs and PUs on the average best achievable capacity assuming a Rayleigh fading model with unit variance.}
 \label{fig::capacity}
 \end{minipage}
 &
 \begin{minipage}{0.5\textwidth}
 \centering
 \includegraphics[width = 2.36 in]{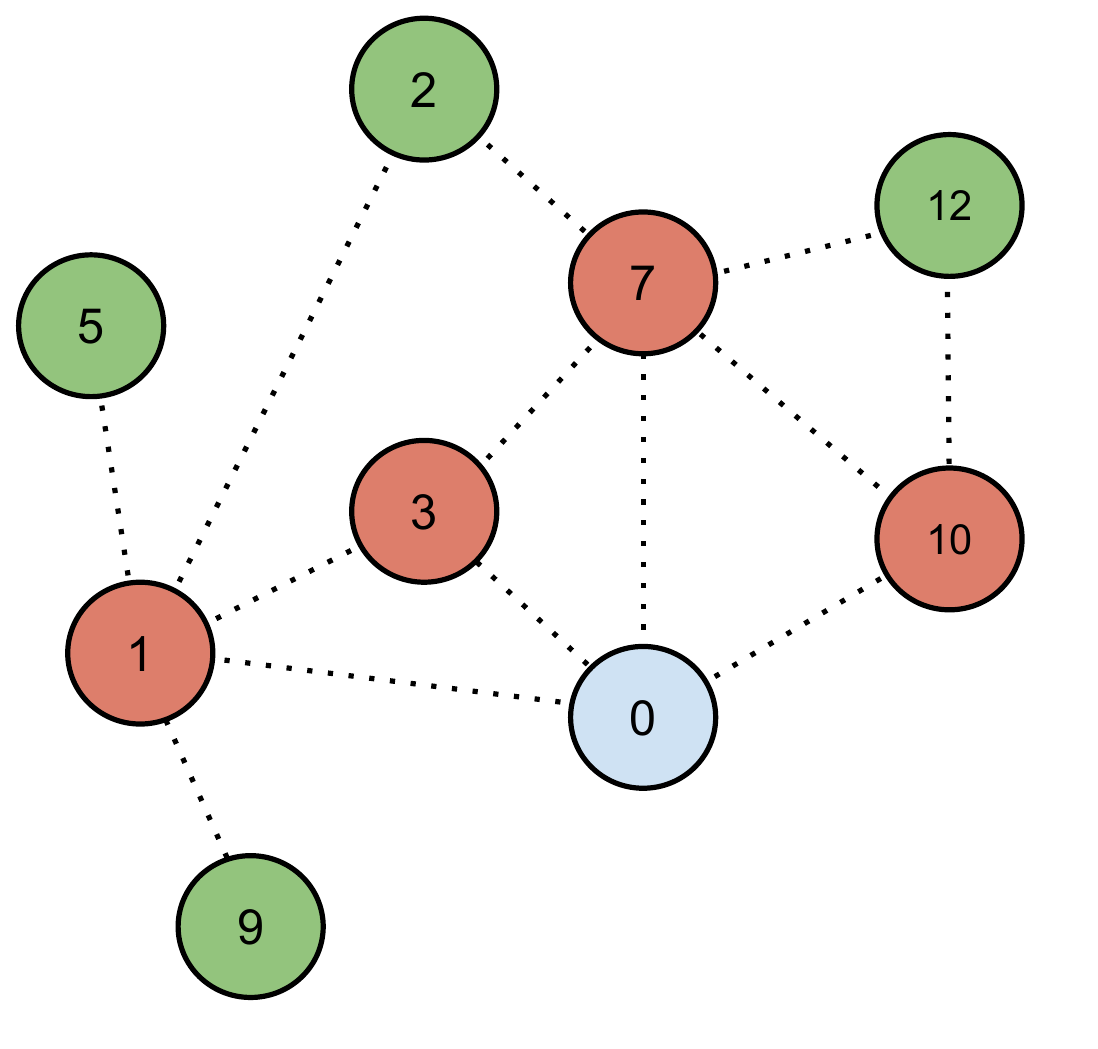}
 \caption{An example on network scenario that shows the effect of choosing different groups on the achievable capactity.}
 \label{fig::neighbors_algorithm}
 \end{minipage}
 
 \end{tabular}
 \end{figure*}

 \subsubsection{Average achievable capacity}
 \label{algorithm}
 In this subsection, we study the statistical behavior of the achievable capacity of different links, and accordingly eliminate cooperation possibilities that are less probable of attaining additional gains in link metric. Considering that the channel coefficients and, consequently, the beamforming weights presented in equations (\ref{capacity_simplex}) and (\ref{capacity_cooperative_group}) are random variables. This renders the achievable capacity of a particular link also of a statistical nature.

In fact, based on the wireless fading model and the instances of the channel coefficients, the achievable capacity of a node or cooperative group links can be foreseen to fall in certain regions with high probability. Conversely, relatively high levels of capacity are not likely to occur given a particular number of nodes in a cooperative group and a given number of active PUs. Figure \ref{fig::capacity} shows the average maximum achievable capacity based on the definitions given in Section \ref{capacity_calculations} for different numbers of cooperating nodes and different numbers of active PUs, for a Rayleigh fading model with unit variance. Note that the achievable capacity in this case is measured by bits per second per Hertz. This value should be multiplied by the used bandwidth to convert it to bits per second (bps). 

For example, considering the case where no PUs are available, it can be seen that moving from a single transmitting node to the case of two cooperating nodes nearly multiplies the average maximum achievable capacity of the link by a factor of 1.5, and moving to higher number of cooperating nodes is accompanied with less relative gains in the achievable capacity. A direct implication follows: considering the link metric construction in (\ref{metric}), the additional gain in the attained capacity by including an extra node in the cooperative group cannot be realized unless the accompanying increase in $N_n$ is of less order.

Consider the scenario in Figure \ref{fig::neighbors_algorithm} in which all the presented nodes carry different data flows. Assume that node 0 calculates the link metric considering all possibilities of cooperation. Based on the previous discussion, using node 1  as a cooperative pair will limit the value of the link metric since it will increase the value of $N_n$ by 3 in addition to the four originally added by node 0 - an increment of 75\% in $N_n$. In contrast, using node 10 as a cooperative pair will not increase $N_n$ beyond 4. Therefore, using node 10 will provide better performance than using node 3. The same argument can be constructed for cooperative groups with a larger number of nodes. It is important to highlight that a node that is not relaying data for any of the existing flows is not affected by the incurred interference and, therefore, is not included in the calculations of $N_n$.

\subsubsection{Elimination algorithm}
 Based on the above, our protocol is devised to reduce the search space by excluding the link constructions that are less probable to score maximum link metrics. Each node estimates a local version of the node density based on the two-hop neighbor information it obtains from periodic hello packets as
 \begin{equation}
 \begin{split}
 D_n &= N/A_N
  \end{split}
 \end{equation}
 \noindent where $N$ is the number of one-hop and two-hop neighbors of the node, and $A_N$ is the area of polygon whose vertices are the farthest of these neighbors from the node. A rough estimate of the number of flows per node $F_n$ can then be calculate mathematically as $F_n = D_f/D_n$. A node can now calculate an estimate of the number of flows that would be affected by the inclusion of $N$ nodes in a group as $N \times F_n$. According to this estimate, and based on the capacity gains shown in Figure \ref{fig::capacity}, a node then decides to exclude group formations that exceed a certain number of collaborating nodes. A node calculates the following factor $\dfrac{N \times F_n - N_{min}}{N_{min}}$ where $N_{min}$ is the value of $N_n$ for the minimum allowable group size (which is the number of PUs plus one as mentioned before in Section \ref{sec::link_formation}). Based on comparing this factor to the corresponding threshold in Table \ref{table_threshold}\footnote{Note that the values given in the table are for particular channel model and statistics (Rayleigh channel model with unit variance), and would differ for other channel conditions. Values of this table depends on plotted values in Figure \ref{fig::capacity}. Providing analytical expressions for the thresholds is a direct extension of this work and is scheduled for later publications.}, the algorithm excludes groups of certain sizes from the search process. We calculate values of Table \ref{table_threshold} as follows: for each row of the table, an entry under group size $j$ is the percentage of increase in the achievable capacity, with corresponds to the achievable capacity of the minimum allowable group size. For example, the entry under group size seven in the second row is $174.1\%$. This means that the capacity achieved by groups of size seven is on average $174.1\%$ greater than the capacity achieved by groups of size three (which is the minimum allowable group size).

 \begin{figure*}
 \centering
  \includegraphics[width = 6.2 in]{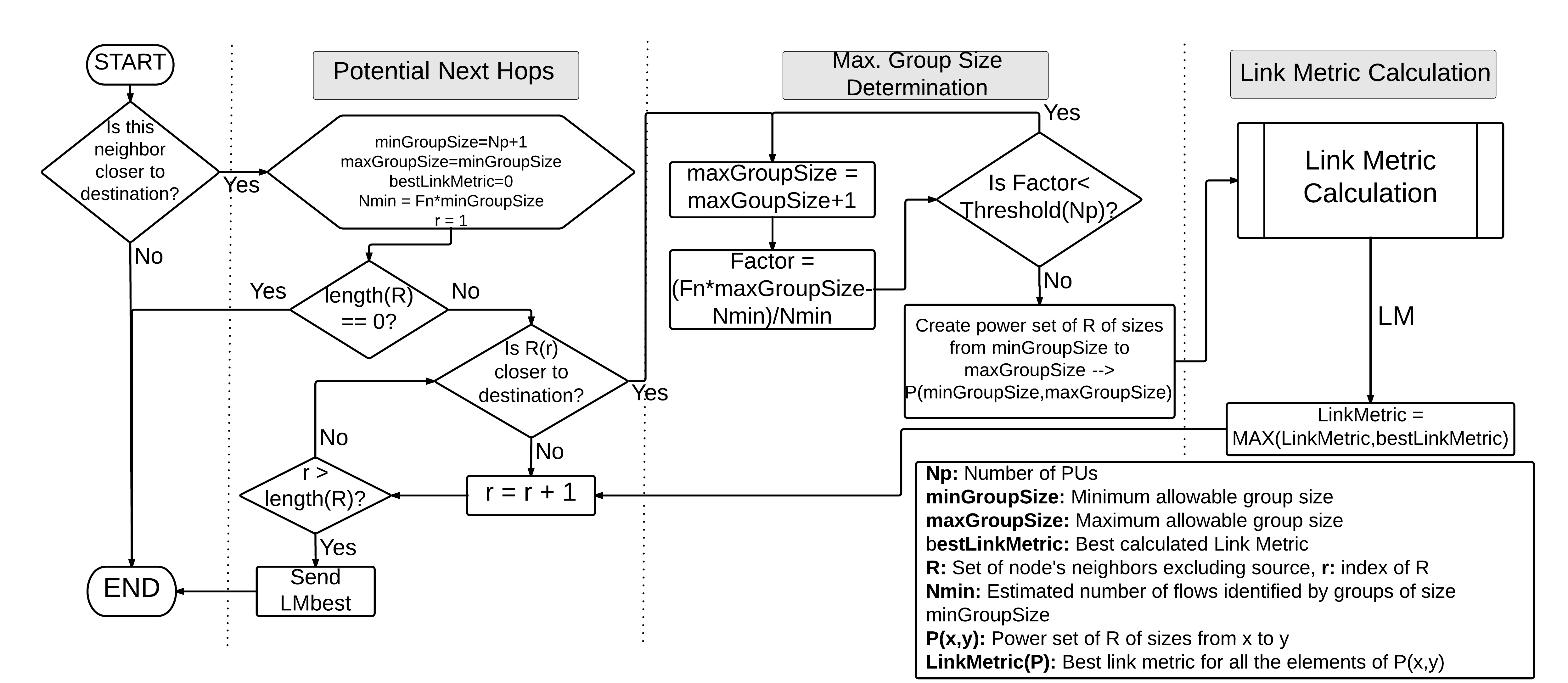}
  \caption{Flowchart of the route reply algorithm applied by each neighbor of the source node.}
  \label{fig::rrep_algorithm}
 \end{figure*}


\subsection{Information Exchange}
In order to allow for the calculation of the achievable capacities across links and the estimation of the on-going flows, a node should be provided periodically with the following information:
\begin{enumerate}
 \item its direct neighbors, and the channel coefficients between the node and each neighbor,
 \item the ID of the on-going flows witnessed by each of the neighbors,
 \item the primary receivers that are detected by each of the neighboring nodes, and the estimated channel coefficients, and
 \item its 2-hop neighbors, and the channel coefficient between each of these neighbors and its intermediate 1-hop neighbor.
\end{enumerate}

We allow nodes to send periodic ``Hello'' packets to their neighbors. A Hello packet consists of: 1) the ID of the generating node, 2) the IDs of the neighboring nodes, along with the channel coefficients between the node and each of its neighbors, 3) the IDs of the identified PUs, along with the estimated channel coefficients from the node to them, and 4) the IDs of the flows witnessed by the node\footnote{This can be obtained by the node by examining the header content of the packets belonging to the flow.}. Given this information, a node can calculate the three mandatory estimates ($D_n$, $D_f$ and $F_n$) as discussed in Section \ref{sec:routeMetric} for link metric calculations.

\begin{figure*}[!t]
\centering
	\begin{subfigure}[t]{0.32\textwidth}
	\centering
    \includegraphics[width=2.3in]{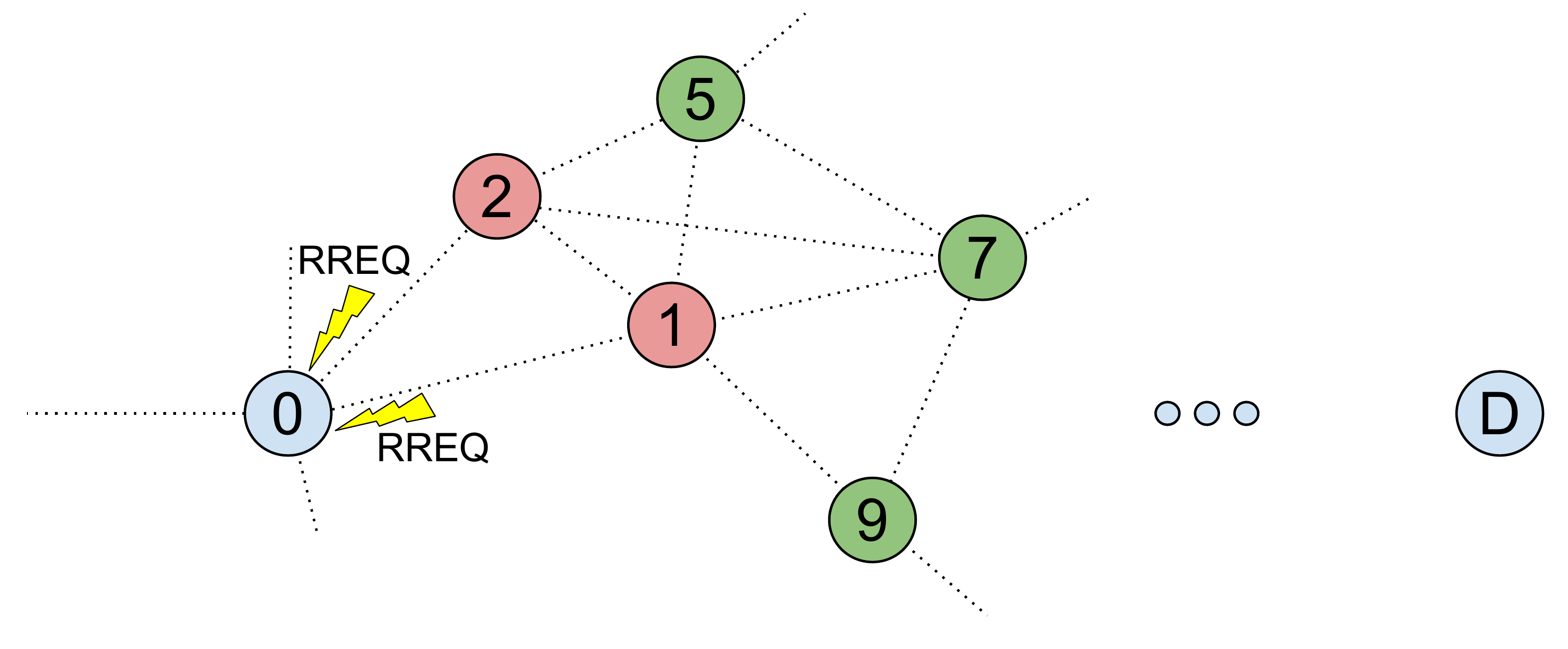}
	\caption{Some node on the route (node 0) sends RREQ to its neighbors.}
	\label{fig::a}
	\end{subfigure}
	\begin{subfigure}[t]{0.32\textwidth}
	\centering
    \includegraphics[width=2.3in]{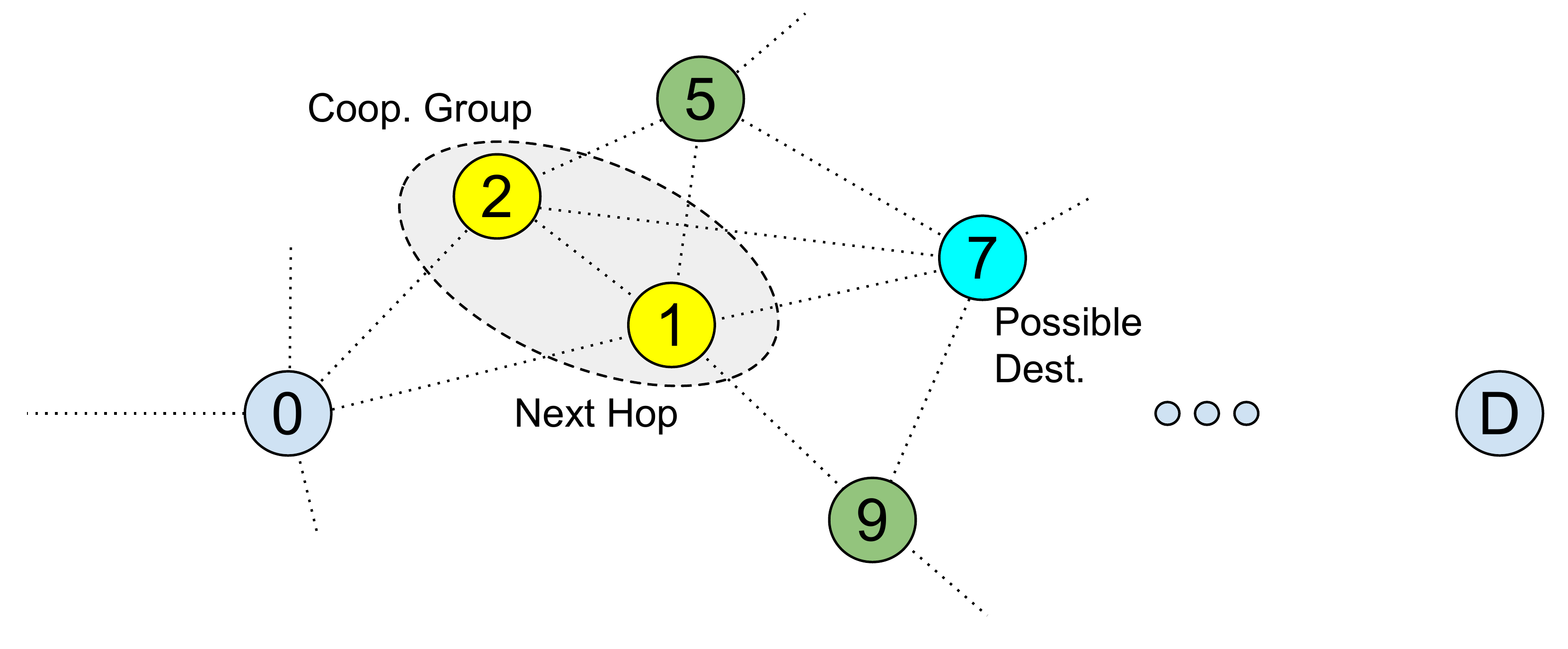}
	\caption{Trying groups of 2 for some possible destination (node 7).}
	\label{fig::b}
	\end{subfigure}
	\begin{subfigure}[t]{0.32\textwidth}
	\centering
    \includegraphics[width=2.3in]{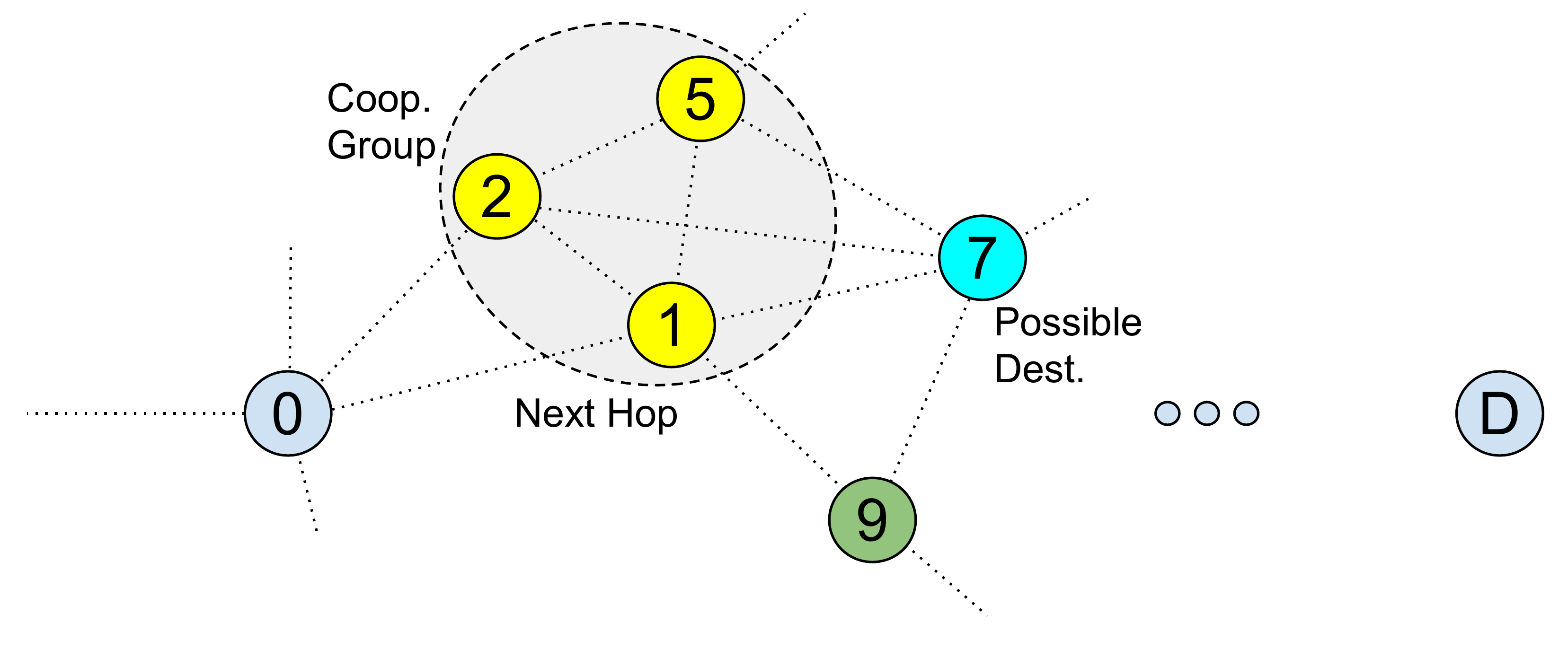}
	\caption{Trying groups of 3 for some possible destination (node 7).}
	\label{fig::c}
	\end{subfigure}
	\begin{subfigure}[t]{0.32\textwidth}
	\centering
    \includegraphics[width=2.3in]{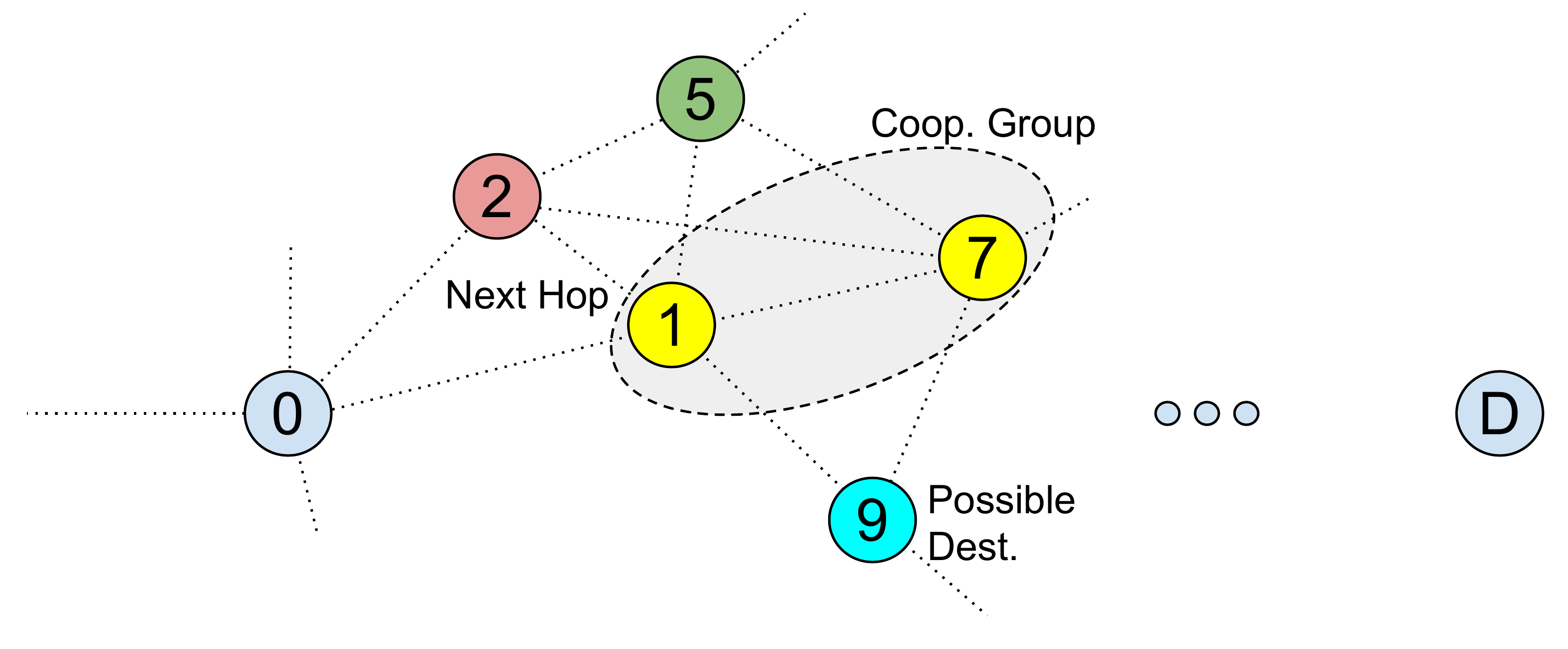}
	\caption{Trying groups of 2 for some possible destination (node 9).}
	\label{fig::d}
	\end{subfigure}
	\begin{subfigure}[t]{0.32\textwidth}
	\centering
    \includegraphics[width=2.3in]{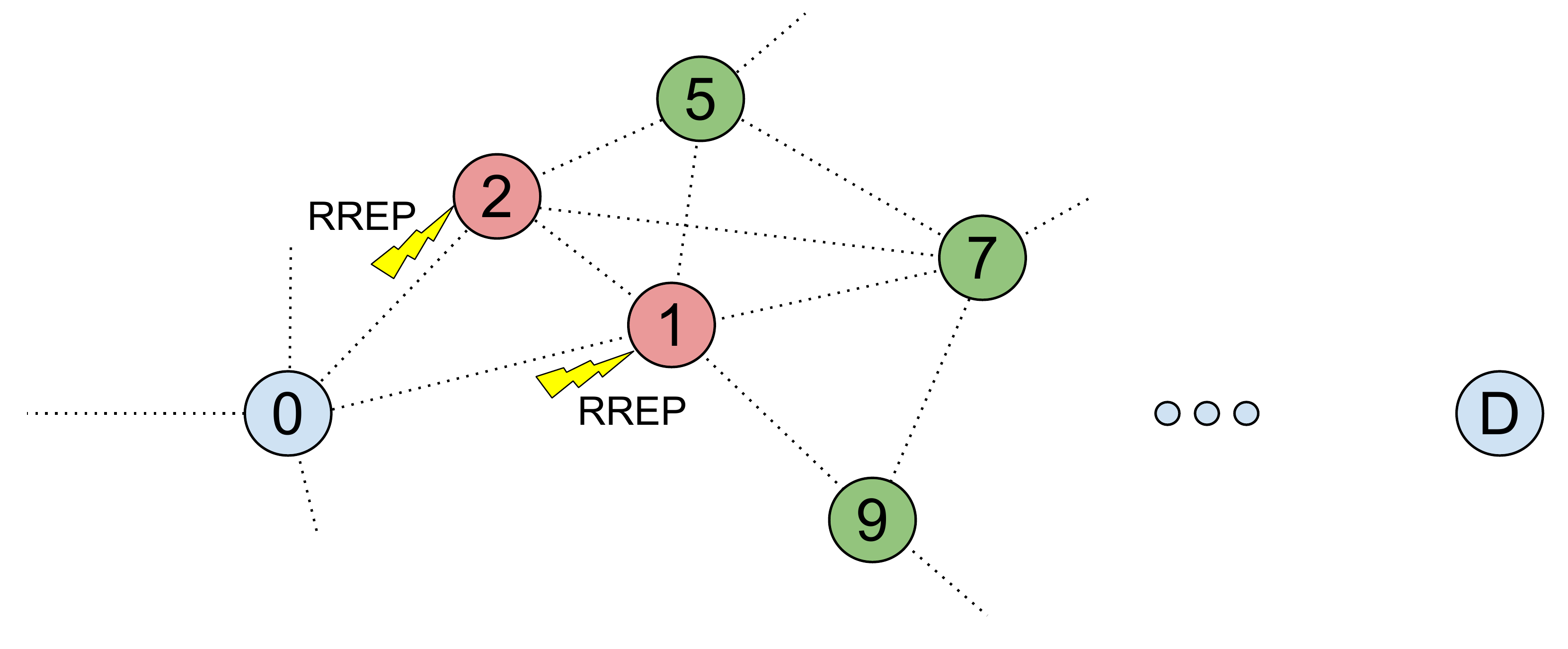}
	\caption{All neighbors send their RREP to the sender node.}
	\label{fig::e}
	\end{subfigure}
	\begin{subfigure}[t]{0.32\textwidth}
	\centering
    \includegraphics[width=2.3in]{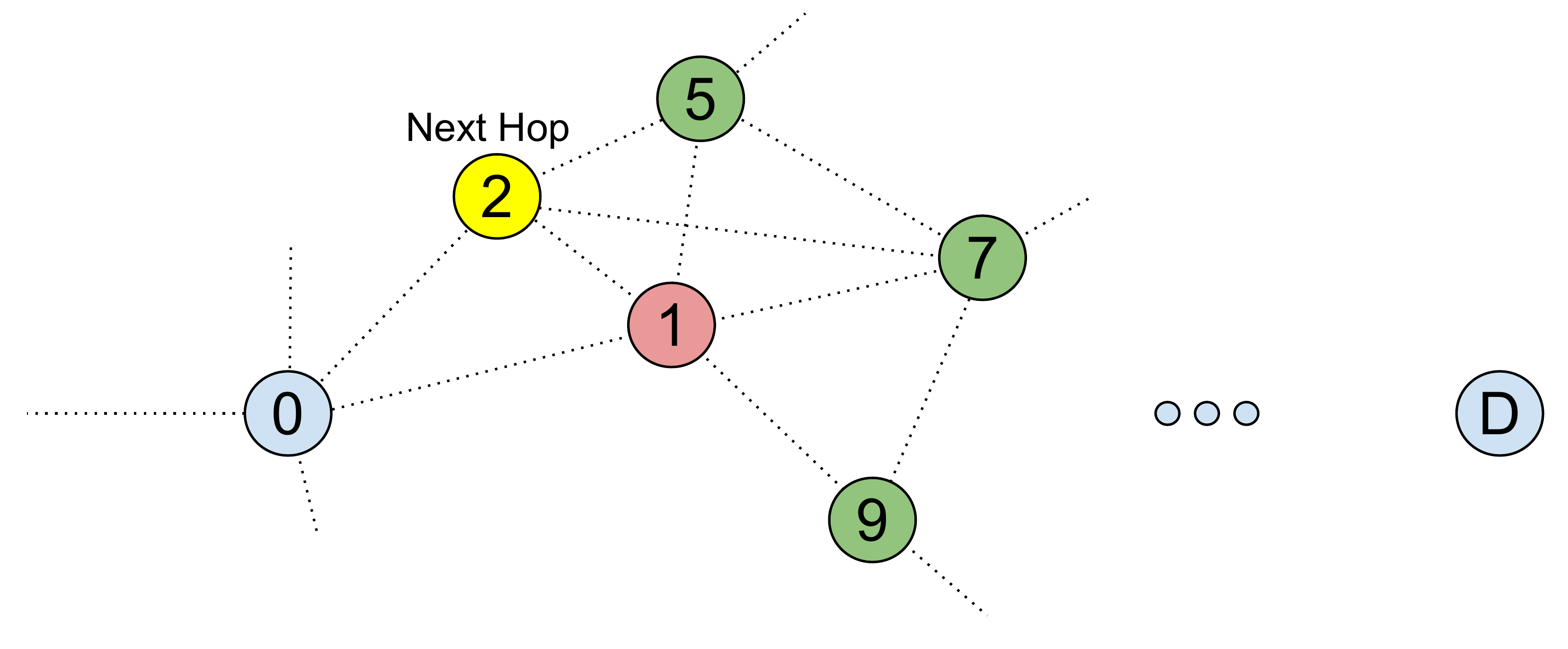}
	\caption{Based on the RREPs, node 0 chooses node 2 to be the next hop.}
	\label{fig::f}
	\end{subfigure}
\caption{\sys{} routing scheme scenario.}
\label{fig::scenario}
\end{figure*}

\subsection{Route Discovery}
A source sends a Route Request (RREQ) packet which contains the IDs of the source and destination nodes. Upon receiving a RREQ, a neighboring node decides whether to send a Route Reply (RREP) packet or to discard the request (if its distance to the destination is greater than that of the source), according to the algorithm described in Figure \ref{fig::rrep_algorithm}. This algorithm is dissected into three phase:
\begin{itemize}
 \item Potential next hops: at the neighbor node, the algorithm sweeps over all possible next hops for the relaying node applying the algorithm. Only next hops that are closer to the destination are considered. First, minimum group size is set to be higher than number of neighboring PUs. Also, the value of $N_{min}$ is calculated in this phase as described in Section \ref{algorithm}. All neighboring nodes are tested one after another for being closer to the destination. If the neighbor is closer to the destination, the algoithm moves to the second phase and if not, this neighbor is discarded and the next neighbor is tested.
 \item Maximum Group Size Determination: basically, maximum group size is set at least one plus the minimum group size. Then, the factor described in Section \ref{algorithm} is calculated. Based on the elimination thresholds given in Table \ref{table_threshold}, the maximum allowable size is determined. At this state, we have the minimum and the maximum groups sizes. We create all possible combinations of groups of all size in the described range for the current potential next hop. Each possible group enters the last stage.
 \item Link Metric Calculation: the algorithm calculates the link metric (Equation \ref{metric}) for all possible transmission configurations between the node applying the algorithm and the possible next hop. If the current metric is the maximum, it is stored in a variable as the maximum metric found for the current node. Finally, The maximum link metric is then sent back to the source node in a RREP.
\end{itemize}

Once a node finishes the route reply algorithm, it sends a RREP packet which consists of 1) the source ID originating the RREQ, 2) the destination ID of the RREQ, and 3) the ID of the next node generating the RREP.

\subsection{Example}
The scenario in Figure \ref{fig::scenario} shows how the routing protocol works. Suppose that some intermediate node (node 0) has  a data packet to forward to some destination (node D). First, it sends a RREQ (Figure \ref{fig::a}) to all its direct neighbors which are nodes 1 and 2. Each of these nodes applies the route reply algorithm to choose the best group to cooperate with. For example, node 1 (for some possible intermediate destination, let it be node 7, as shown in Figure \ref{fig::b}) tries to construct group of two nodes with all of its direct neighbor (other than the sender node). Then, it tries to construct a group of three nodes (Figure \ref{fig::c}) and so on. It repeats the same scenario for all possible intermediate destinations like node 9 in Figure \ref{fig::d}. After finishing, all neighbors of the source node send back their RREPs (Figure \ref{fig::e}). Finally, the sender node chooses one of those neighbors to be the next hop (Figure \ref{fig::f}) based on the routing protocol metric described in Section \ref{sec:routeMetric}. Then the routing process continues to the destination in the same way.


\section{Performance Evaluation} \label{sec::eval}
\begin{table}[!t]
    \centering
    \begin{tabular}{|l |l |l|}
    \hline
    Parameter & Value range & Nominal Value(s)\\
    \hline
    Number of PUs & 2 - 16 & 4\\
    Number of SUs & 10 - 40 & 25\\
    SU transmission range (m) & 125 & 125\\
    PU transmission range (m) & 140 & 140\\
    Number of connections & 1 - 16 & 8\\
    Frequency (GHz) & 2.4 & 2.4 \\
    Effective bandwidth (Mbps) & 1.5 & 1.5 \\
    Packet Size (byte) & 128 - 1518 & 512\\
    PU Activity(\%) &   0 - 100 & 20 \\
    Data Rate Per Source (kbps) & 20 - 400 & 100 \\
    Deployment Area Side Length (m) & 250 - 1000 & 250 \\
    $\tau$ (sec) & 1 & 1 \\
    \hline
    \end{tabular}
    \caption{Experiments parameters.}
    \label{fig::simulation_parameters_summary}
\end{table}

In this section, we evaluate the performance of our proposed routing protocol using a cognitive extension of NS2 \cite{NS2},\cite{crextension}. Table \ref{fig::simulation_parameters_summary} summarizes the simulation parameters used in our evaluation. We model the PUs activity as an ON-OFF process where the means of the exponentially distributed active and inactive periods are randomly chosen (according to a uniform distribution) with the activity percentage shown in Table \ref{fig::simulation_parameters_summary}. PUs are uniformly distributed over the available grid. We assume the channel coefficients to be complex numbers that follow Gaussian distribution with zero mean and unit variance \cite{gomadam2008approaching,yu2002models}.We assume that the SUs are randomly deployed using a uniform distribution across the grid. Each SU node is equipped with two radio interfaces and has omni-directional antennas and runs the IEEE 802.11 MAC protocol. The first radio is used for exchanging the control packets while the second is used for exchanging data. The source and destination of each connection are selected randomly. We compare our protocol against existing protocols such as LAUNCH \cite{habak2013location} and CAODV \cite{caodv}. LAUNCH is a location-aided routing protocol that is designed to work in Cognitive Radio Networks. CAODV is the cognitive extension of the popular AODV protocol \cite{Perkins:2003:AHO:RFC3561}. We have chosen these two protocols as representatives for the local and global approaches of routing protocols of CRNs respectively.

\subsection{Metrics}
We evaluate \sys{} using the following metrics:
\begin{enumerate}
\item Goodput: number of bits communicated successfully from the source to the destination per second.
\item Average end-to-end delay: average time taken by packets to reach the destination from the source.
\item  Routing overhead: number of transmitted control packets in the routing phase.
\item Average group size: average number of nodes participating in the cooperative commuication in case of using \sys{}.
\item Routing Opportunities Gain: average number of groups such a node can construct to route through. This number converges to one if no groups are used and hence we called it gain.
\end{enumerate}

\subsection{Experimental Results}
\subsubsection{Changing Network Density}

\begin{figure}[!t]
\centering
	\begin{subfigure}[t]{0.23\textwidth}
	\centering
    \includegraphics[width=1.7in]{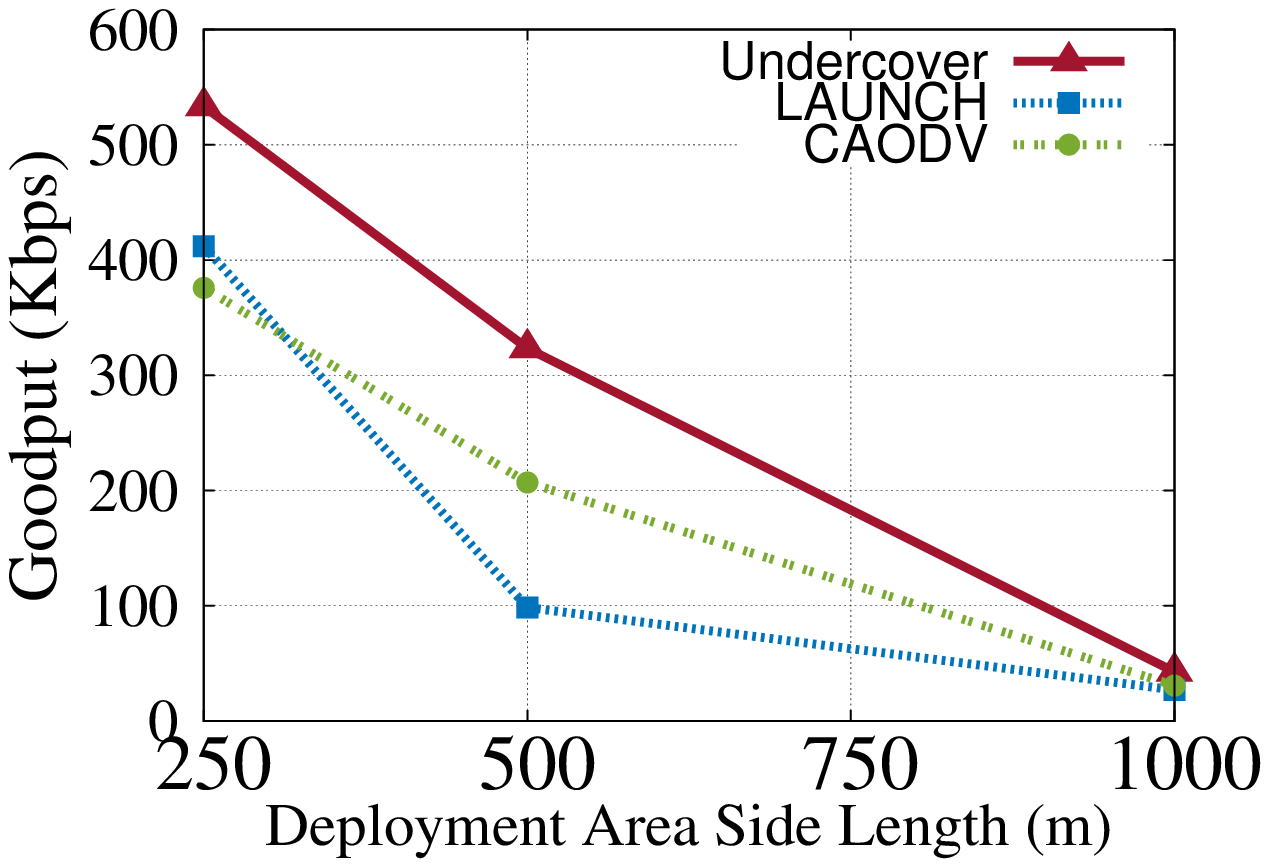}
	\caption{Goodput}
	\label{fig::areathrough}
	\end{subfigure}
	\begin{subfigure}[t]{0.23\textwidth}
	\centering
    \includegraphics[width=1.7in]{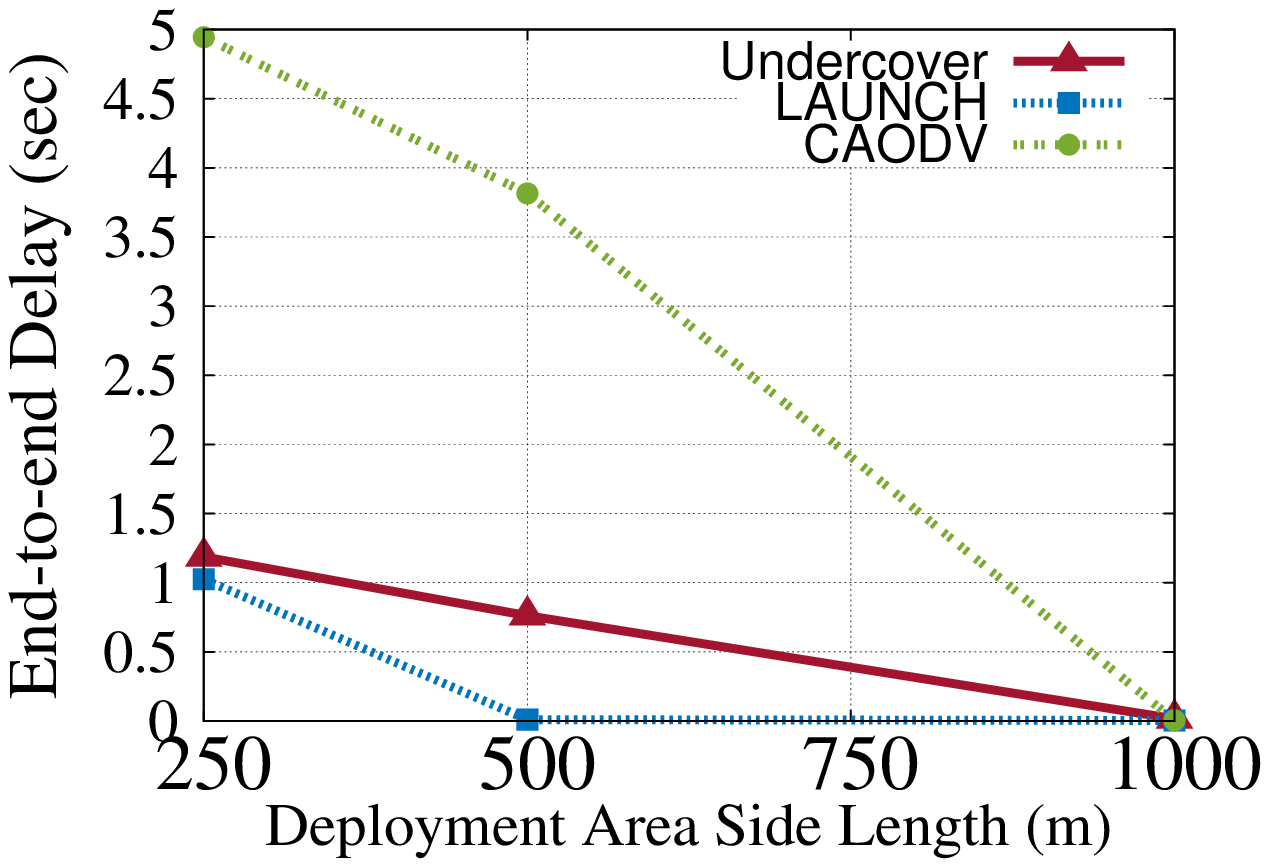}
	\caption{Average End-to-end Delay}
	\label{fig::areadelay}
	\end{subfigure}
\caption{Effect of changing deployment area size on network performance.}
\label{fig::area}
\end{figure}

\paragraph{Changing Deployment Area Size}
Figure \ref{fig::area} shows the effect of changing the deployment area size on the performance metrics. Increasing the square deployment area side length increases its area and decreases the SUs density. We can see the same observations highlighted in the previous section in terms of the behavior of goodput and delay when network density changes. However, it is vital to mention here that using the small area (250m$\times$250m) for all of the next experiments was driven by the need to test \sys{} in dense environment to allow for groups formation.

\begin{figure}[!t]
\centering
	\begin{subfigure}[t]{0.23\textwidth}
	\centering
    \includegraphics[width=1.7in]{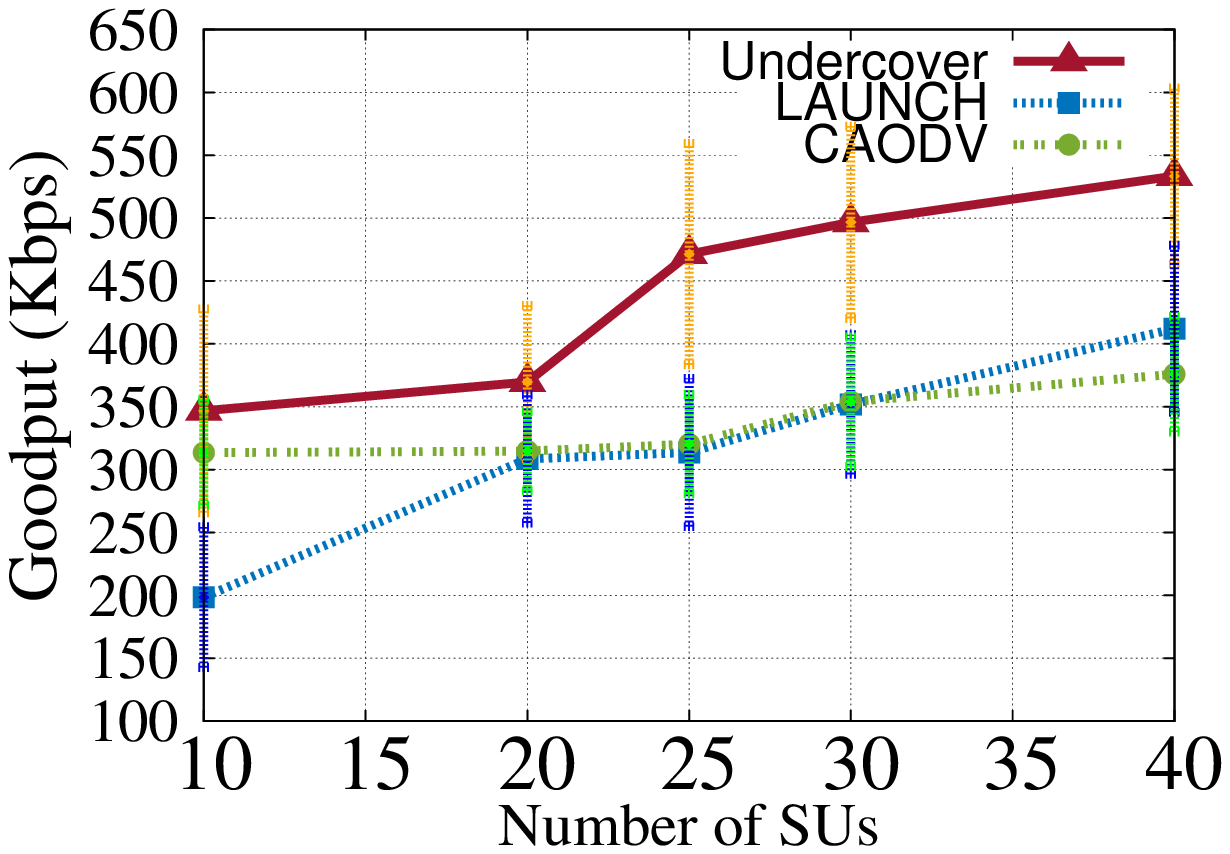}
	\caption{Goodput}
	\label{fig::nsusthrough}
	\end{subfigure}
	\begin{subfigure}[t]{0.23\textwidth}
	\centering
    \includegraphics[width=1.7in]{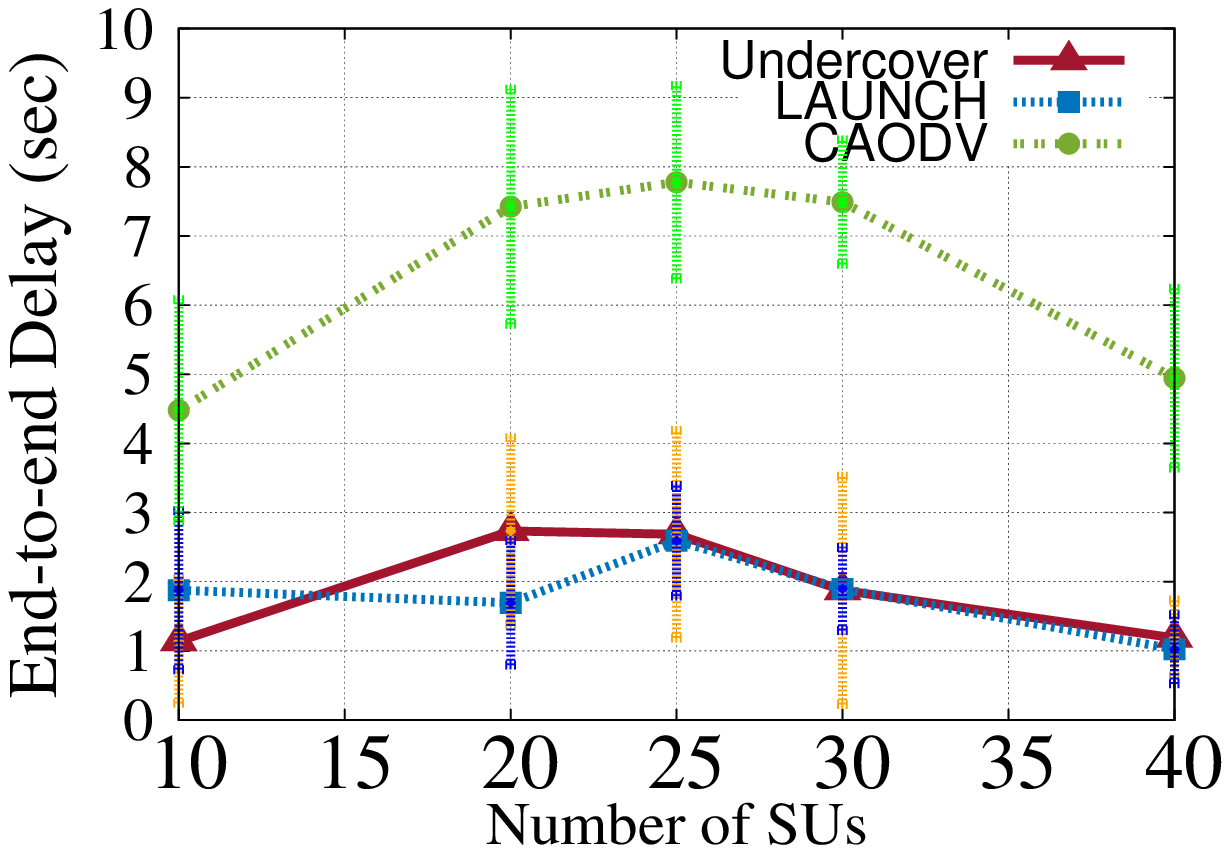}
	\caption{Average End-to-end Delay}
	\label{fig::nsusdelay}
	\end{subfigure}
 	\begin{subfigure}[t]{0.23\textwidth}
 	\centering
  \includegraphics[width=1.7in]{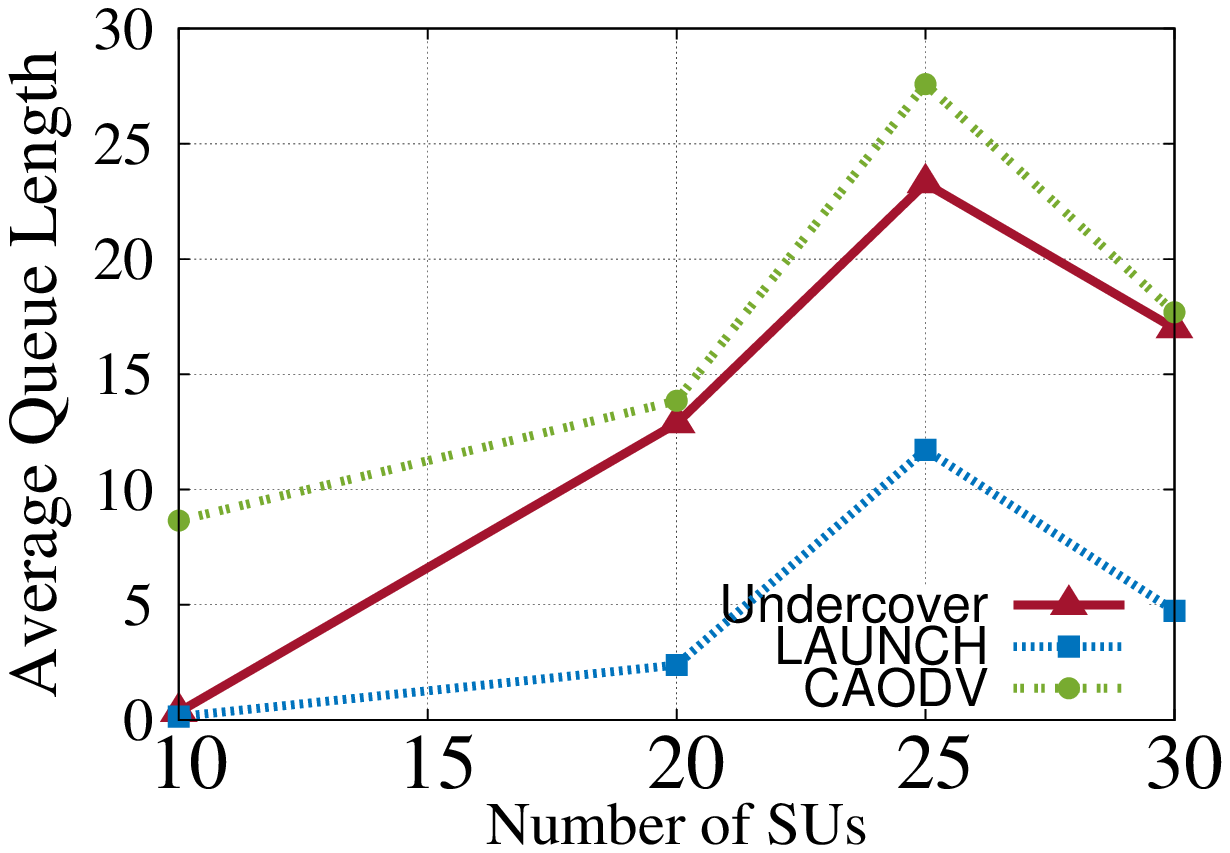}
  	\caption{Average Queue Length.}
  	\label{fig::qlength}
 	\end{subfigure}
	 \begin{subfigure}[t]{0.23\textwidth}
 	\centering
  	\includegraphics[width=1.7in]{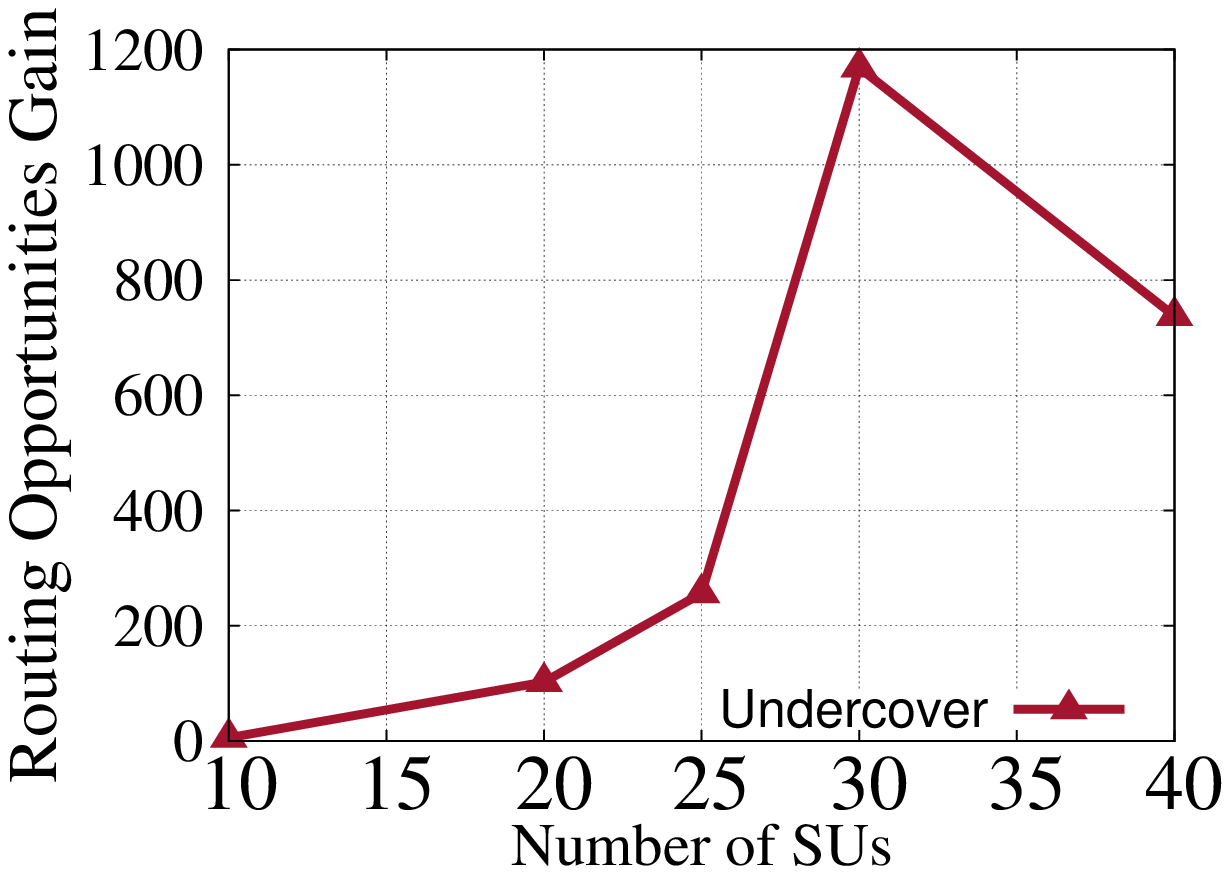}
  	\caption{Routing Opportunites Gain.}
  	\label{fig::routeopp}
 	\end{subfigure}
\caption{Effect of changing number of SUs on network performance.}
\label{fig::nsus}
\end{figure}

\paragraph{Changing Number of SUs}
Figure \ref{fig::nsus} shows the advantage of using \sys{} over CAODV and LAUNCH in terms of the achieved goodput and  average end-to-end delay as the SU density increases. Several conclusions can be drawn from this figure. Generally, goodput increases with the increase of SUs' density. This happens since we can find better routes as the number of SUs increases. Also, we can see that \sys{} outperforms both of  LAUNCH and CAODV in terms of goodput especially at high density of SUs. This is due to the ability of \sys{} to construct better and larger cooperative groups with this increase where, groups of size 8 were attained in some experiments. This leads to more reliable delivery of packets to destinations. Our last note for Figure \ref{fig::nsusthrough} is that in high density networks, LAUNCH beats CAODV since the former technique for routing takes into consideration the minimum delay and the PU presence.

Concerning Figure \ref{fig::nsusdelay}, we can see a bell-like shape with a peak at some point in the graph for all protocols. This behavior can also be observed in Figure \ref{fig::qlength} and can be attributed to the following reason. There are two competing factors that affect the queues length and hence the delay. The first one is the number of transmissions between the sender and the receiver which affects the queuing delay at each node on the route. This factor inceases with the increase of number of SUs as shown in Figure \ref{fig::nsusthrough}. The second one is the advantage of finding better routes as the number of SUs increases which decreases the total end-to-end delay since the probability to interfere with a PU decreases. At the first part of the graph, the first factor beats the second one. Thus, increasing SUs density increases the backoff delay and the number of retransmissions due to the congestion at the MAC layer. This increases the queue length at each node (Figure \ref{fig::qlength}) increasing the total end-to-end delay. However, in the second half of the graph, the opposite case happens where we can see the effect of the second factor. Then, the advantage of finding better routes (and hence getting better experience for data delivery) shows upper hand over the counter effect of increasing queuing delay (due to SUs number increase) i.e. queues length decreases at each node as shown in Figure \ref{fig::qlength}. This leads to the decrease of the average end-to-end delay as the SUs density increases at the end.

\begin{figure}[!t]
 \centering
  \includegraphics[width=0.4\textwidth]{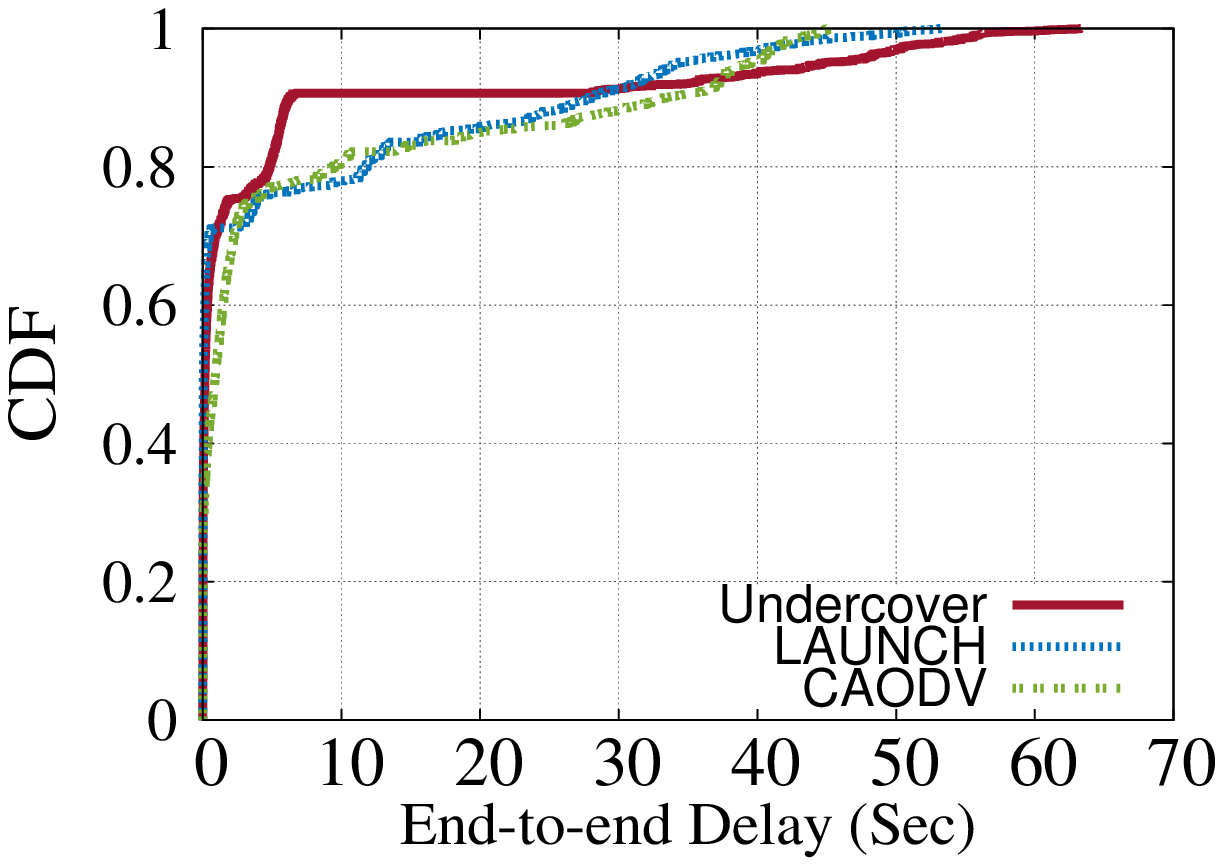}
  \caption{CDF of end-to-end delay when using the default parameters of table \ref{fig::simulation_parameters_summary}.}
  \label{fig::delaycdf}
 \end{figure}

We can see that CAODV always has higher delay than \sys{} and LAUNCH. This happens since the last two protocols have the ability to deal better with the presence of PUs either by constructing cooperative groups or by the channel switching used by LAUNCH. On the other hand, the experienced delays for \sys{} and LAUNCH are nearly equal in almost all cases since both try to avoid interfering PUs by sending on different channels or by nulling tranmission at them. Although, the average end-to-end delay of \sys{} seems to be higher than that of LAUNCH in some cases, we can get from Figure \ref{fig::delaycdf} a more detailed message. We can see that in case of using \sys{}, a very small portion of the packets suffers from excessive delays introduced by route construction and this route is stable enough for more than 90\% of the packets. On the other hand, we can see that less than 80\% of packets transmitted using LAUNCH have the same small delay. Thus, we can conclude that most of packets routed using \sys{} incur very small delay compared to LAUNCH even if it seems to have higher average end-to-end delay. Another conclusion can be driven from that figure: it seems that the groups construction phase takes a lot of time which makes about 10\% on average have a high delay. This fact opens for us a room of improvements which can be done as a future extension on the current work. This conclusion can be applied too to the rest of the average end-to-end delay figures in this section.

Figure \ref{fig::routeopp} shows the effect of changing number of SUs on the routing opportunities gain. This figure mainly compares between \sys{} and LAUNCH since the former protocol converges to LAUNCH in case of not using groups. Thus, this figure shows the advantage of using cooperation with neighboring nodes. We can see that using cooperative groups gives more routing opportunities in all cases. This leads to the discovery of better and more stable paths and this fact illuminate the value of cooperation used by \sys{}. We can see that opportuniteis increase with the increase of number of SUs, since more nodes exist to cooperate with. However, at very dense network (when number of SUs = 40), the number of opportunities decreases. This happens since not all the potential groups are taken into consideration due to the criteria we put in Section \ref{sec::link_formation} to shrink the search space. This helps us to save the computation time at each node. On the other hand, number of opportunities offered by the non-cooperative protocol increases. Both factors lead to the decrease of the opportunities gain at the end of the figure. 

\begin{figure}[!t]
\centering
	\begin{subfigure}[t]{0.23\textwidth}
	\centering
    \includegraphics[width=1.7in]{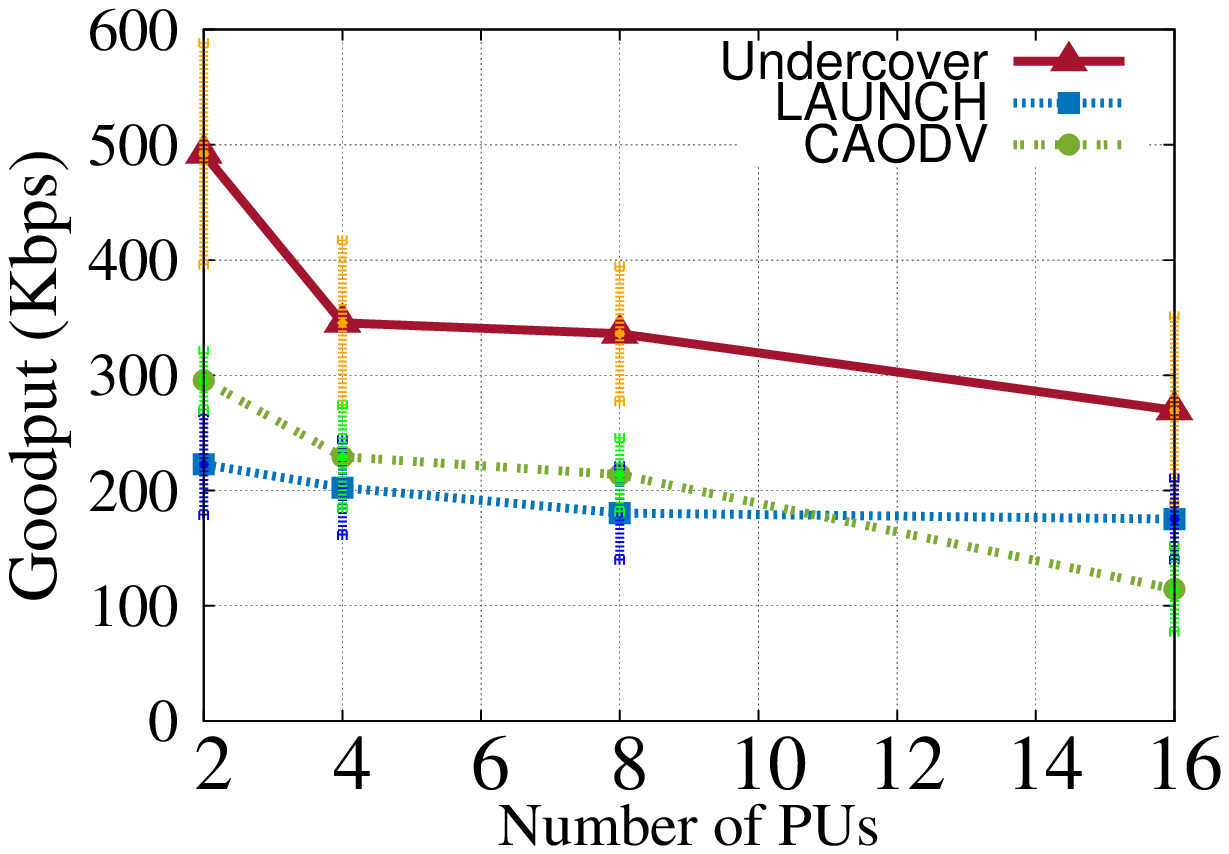}
	\caption{Goodput}
	\label{fig::npusthrough}
	\end{subfigure}
	\begin{subfigure}[t]{0.23\textwidth}
	\centering
    \includegraphics[width=1.7in]{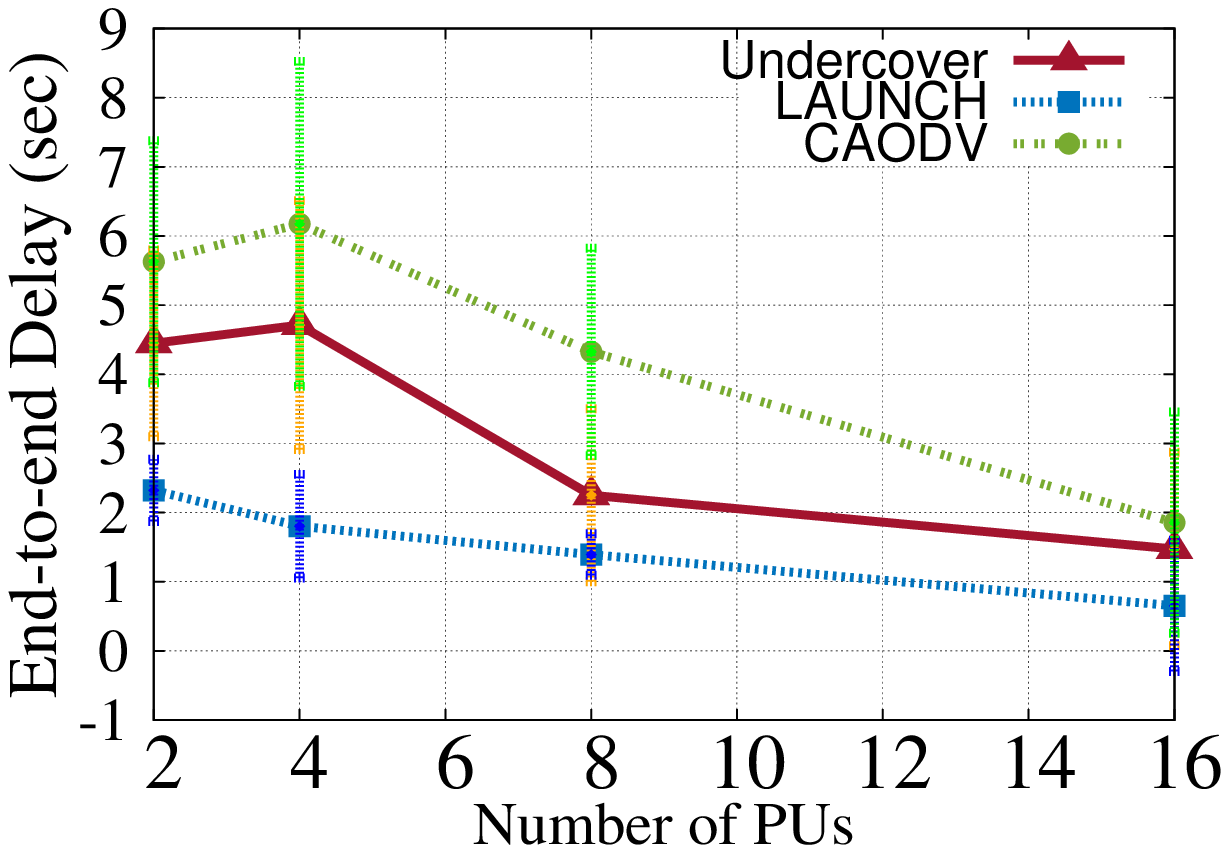}
	\caption{Average End-to-end Delay}
	\label{fig::npusdelay}
	\end{subfigure}
\caption{Effect of changing number of PUs on network performance.}
\label{fig::npus}
\end{figure}

\subsubsection{Changing Number of PUs} 
Figure \ref{fig::npus} shows the effect of changing number of PUs on the goodput and the end-to-end delay for the three used protocols for comparison. It can be noted that the performance for the three protocols degrades as the number of PUs increases. However, the performance of \sys{} always outperforms that of LAUNCH and CAODV in terms of the achieved goodput (Figure \ref{fig::npusthrough}). This happens since this target is our main concern while designing our new protocol. Thanks to the constructed cooperative group, \sys{} can send in many cases without interfering PUs even if they exist and active. However, the overhead of creating these groups in terms of elapsed time and interference to others can be shown in Figure \ref{fig::npusdelay} in which we can see that the delay for \sys{} is higher than that of LAUNCH. We can note also in this figure that generally, the delay decreases as number of PUs increases. Due to the greater effect of PUs as their number increases, the packet loss ratio increases. Thus, the number of correctly delivered packets to their destinations decreases, the queues length at each node decreases, and hence decreasing the end-to-end delay.

\begin{figure}[!t]
\centering
	\begin{subfigure}[t]{0.23\textwidth}
	\centering
    \includegraphics[width=1.7in]{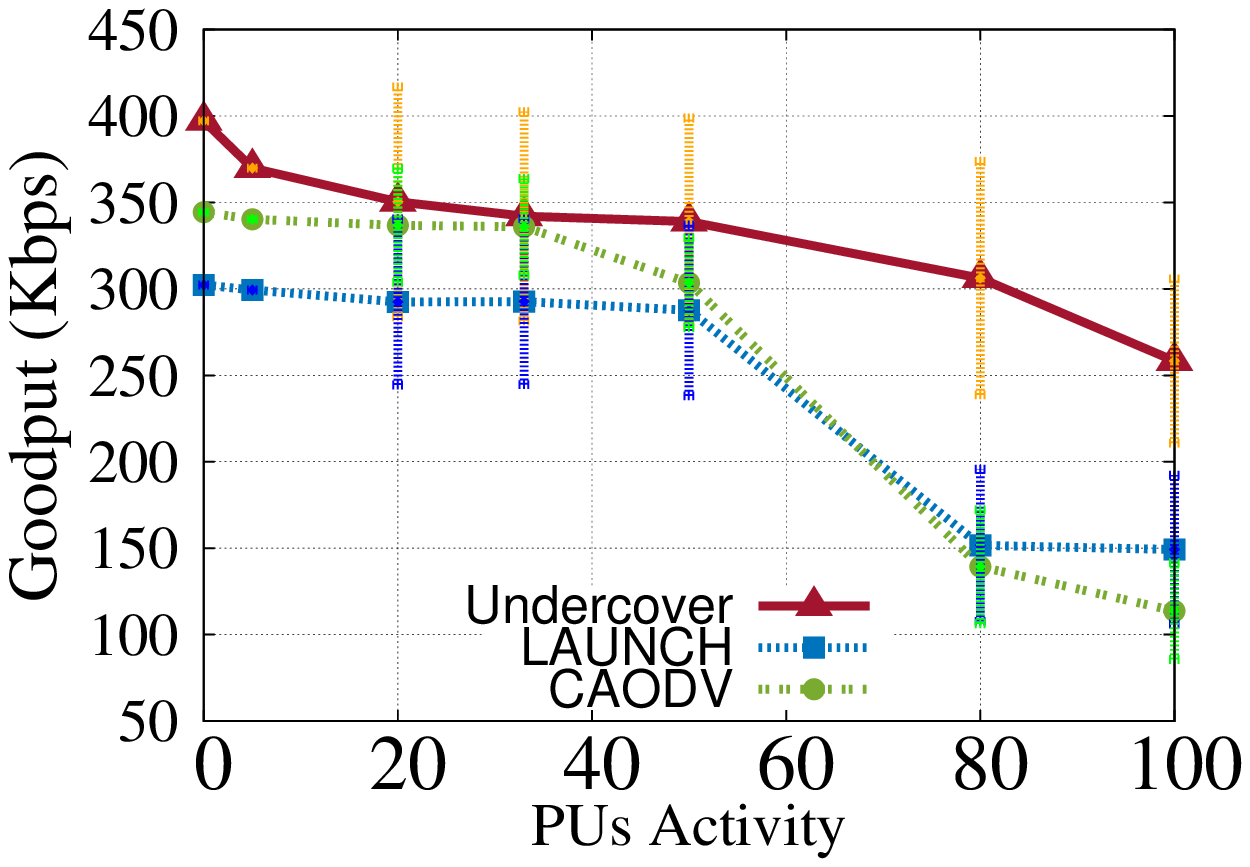}
	\caption{Goodput}
	\label{fig::puactivitythrough}
	\end{subfigure}
	\begin{subfigure}[t]{0.23\textwidth}
	\centering
    \includegraphics[width=1.7in]{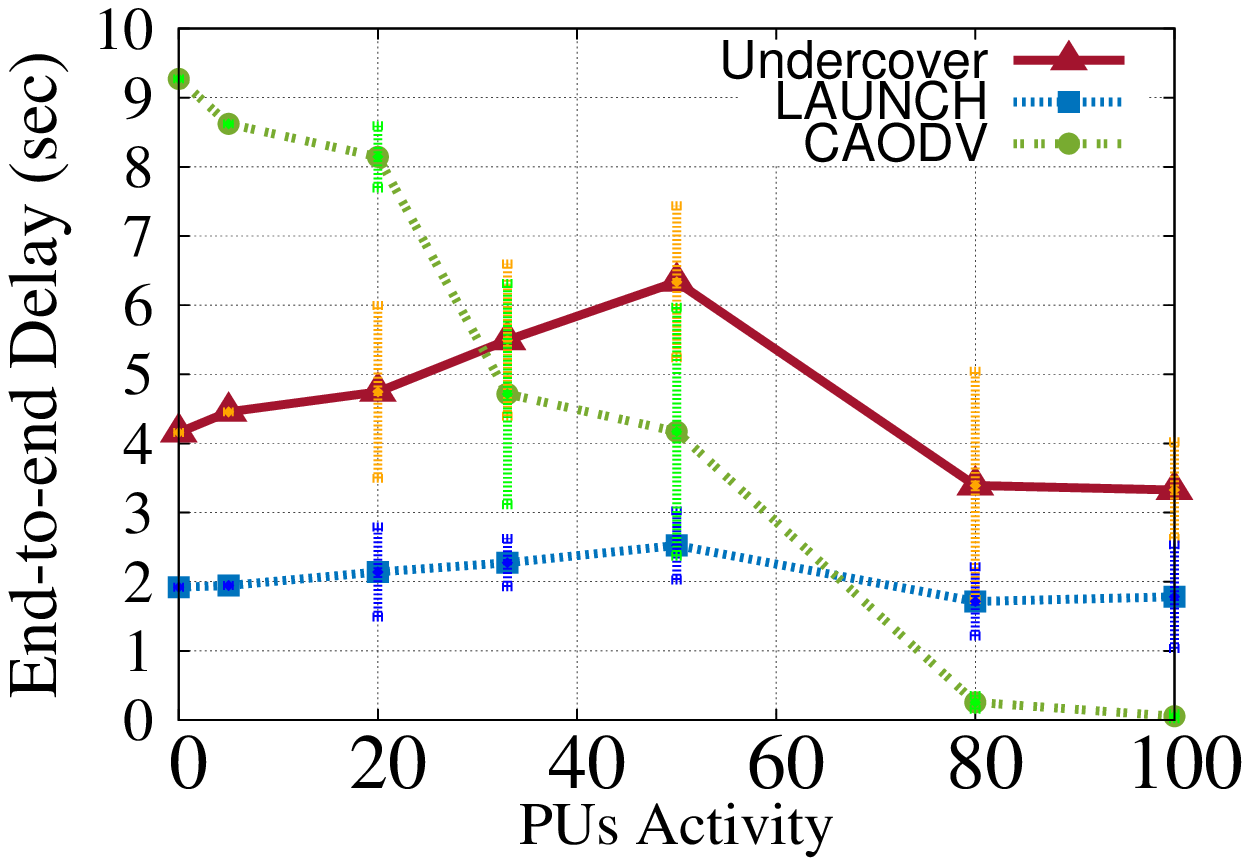}
	\caption{Average End-to-end Delay}
	\label{fig::puactivitydelay}
	\end{subfigure}
\caption{Effect of changing PUs activity on network performance.}
\label{fig::puactivity}
\end{figure}
 
\subsubsection{Changing PUs Activity} 
Figure \ref{fig::puactivity} shows the effect of changing PUs activity on the performance metrics. We can conclude from Figure \ref{fig::puactivitythrough} that the goodput decreases with the increase of PUs activity. This happens since less number of packets are able to reach their destinations safely with the increase of PUs activity decreasing the goodput with always upper hand for \sys{} over other protocols. There is also an important observation in this figure. We can see that \sys{} beats other protocols in case of zero activity of PUs. In this case, the network is converted from Cognitive to Adhoc network i.e. no PU traffic in this case. This means that CAODV converges to AODV and LAUNCH advantage in cognitive networks disappears. Cooperative groups constructed by \sys{} in this case gives source node the ability to send with higher rate and hence achieving higher goodput. Thus, this figure shows the second goal of constructing cooperative groups which is strengthing the sending power to get a signal with better quality. We can note some other facts from Figure \ref{fig::puactivitydelay}. Generally, the average end-to-end delay decreases with the increase of PUs activity. This happens due to delivering less number of packets and hence decreasing the congestion at MAC layer and the queue length at each node decreasing the total delay. We can note that at 100\% activity, CAODV achieves the lowest time delay and the lowest goodput (highest loss rate). However, the delay increases at the first part of the graph for some protocols due to the time spent in constructing cooperative groups (to overcome PUs existence and activity) in case of \sys{} and the channel switching done by LAUNCH. But at high rates of PUs, the delay decreases finally for all protocols.

\begin{figure}[!t]
\centering
	\begin{subfigure}[t]{0.23\textwidth}
	\centering
    \includegraphics[width=1.7in]{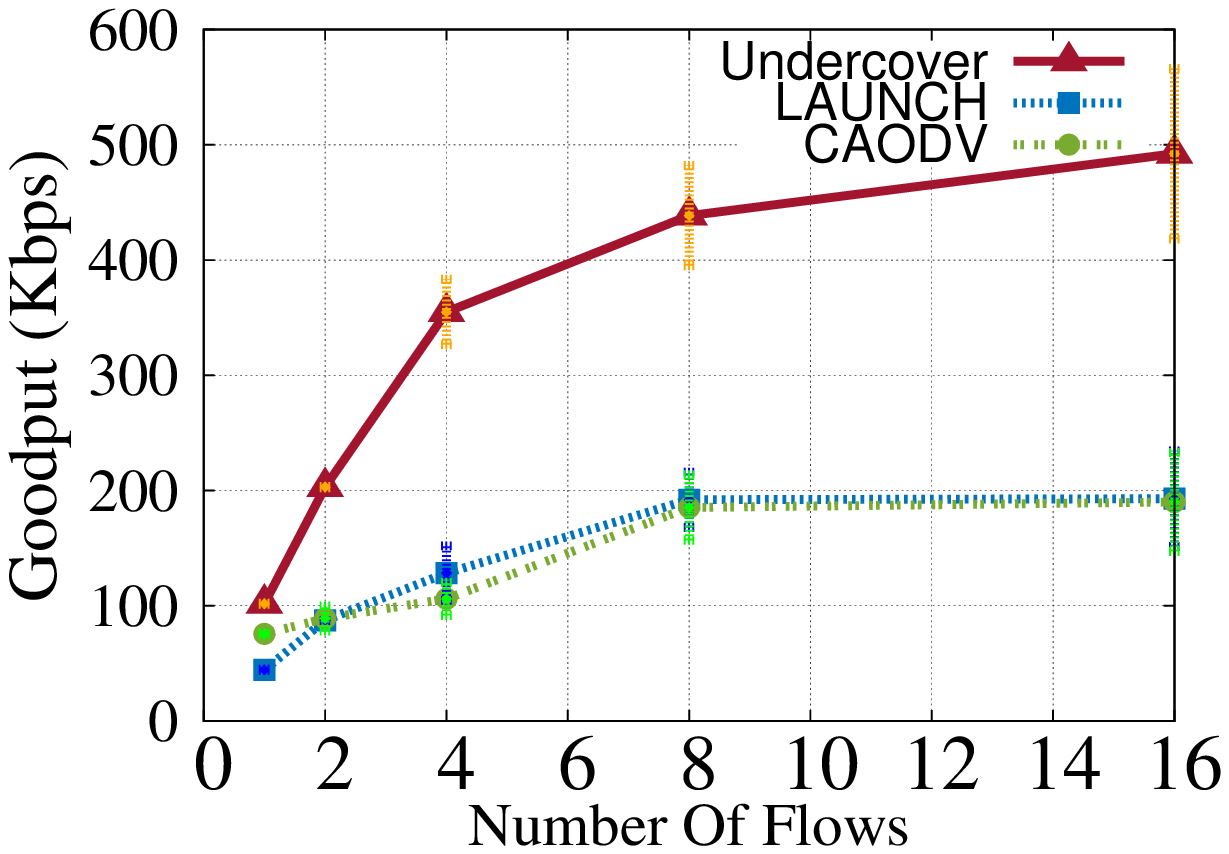}
	\caption{Goodput}
	\label{fig::nflowsthrough}
	\end{subfigure}
	\begin{subfigure}[t]{0.23\textwidth}
	\centering
    \includegraphics[width=1.7in]{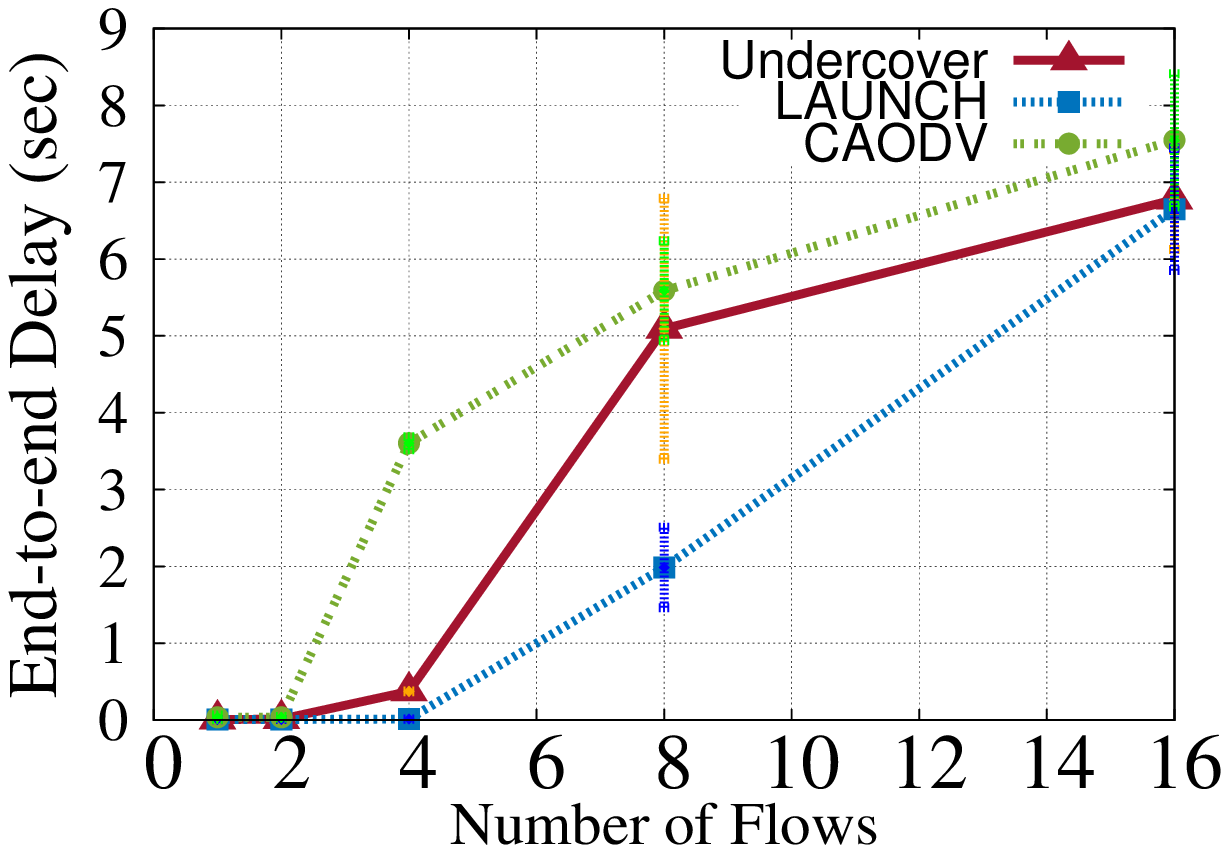}
	\caption{Average End-to-end Delay}
	\label{fig::nflowsdelay}
	\end{subfigure}
\caption{Effect of changing number of flows on network performance.}
\label{fig::nflows}
\end{figure}
 
\subsubsection{Changing Number of Flows}
Figure \ref{fig::nflows} shows the effect of increasing the number of active connections on the performance metrics. From Figure \ref{fig::nflowsthrough}, we can note that \sys{} has the best performance when there is a large number of flows in the network. Where goodput saturates in case of using LAUNCH and CAODV while still increasing (even with a decreased rate) in case of using \sys{}. This happens in case of LAUNCH and CAODV since there are not enough SUs to accomodate flows. The same saturation effect occurs with \sys{} but at a later point on the graph. This can be abstractly explained by thinking of networks operating with \sys{} as larger networks with virtual SUs that correspond to groups. Also, increasing number of flows increases the goodput as well. This is due to the fact of transferring more packets to destinations which increases the goodput value by definition. We can see that \sys{} performance exceeds that of CAODV and LAUNCH due to the ability of delivering more packets for each single added flow, which is translated to enhanced performance. From Figure \ref{fig::nflowsdelay}, we can see an increase in the values of delay with the increase of the number of active connections. This is due to the increase of the number of deliverd packets to their destinations and hence, the increase of the delay due to congestion at the MAC layer.
\begin{figure}[!t]
\centering
	\begin{subfigure}[t]{0.23\textwidth}
	\centering
    \includegraphics[width=1.7in]{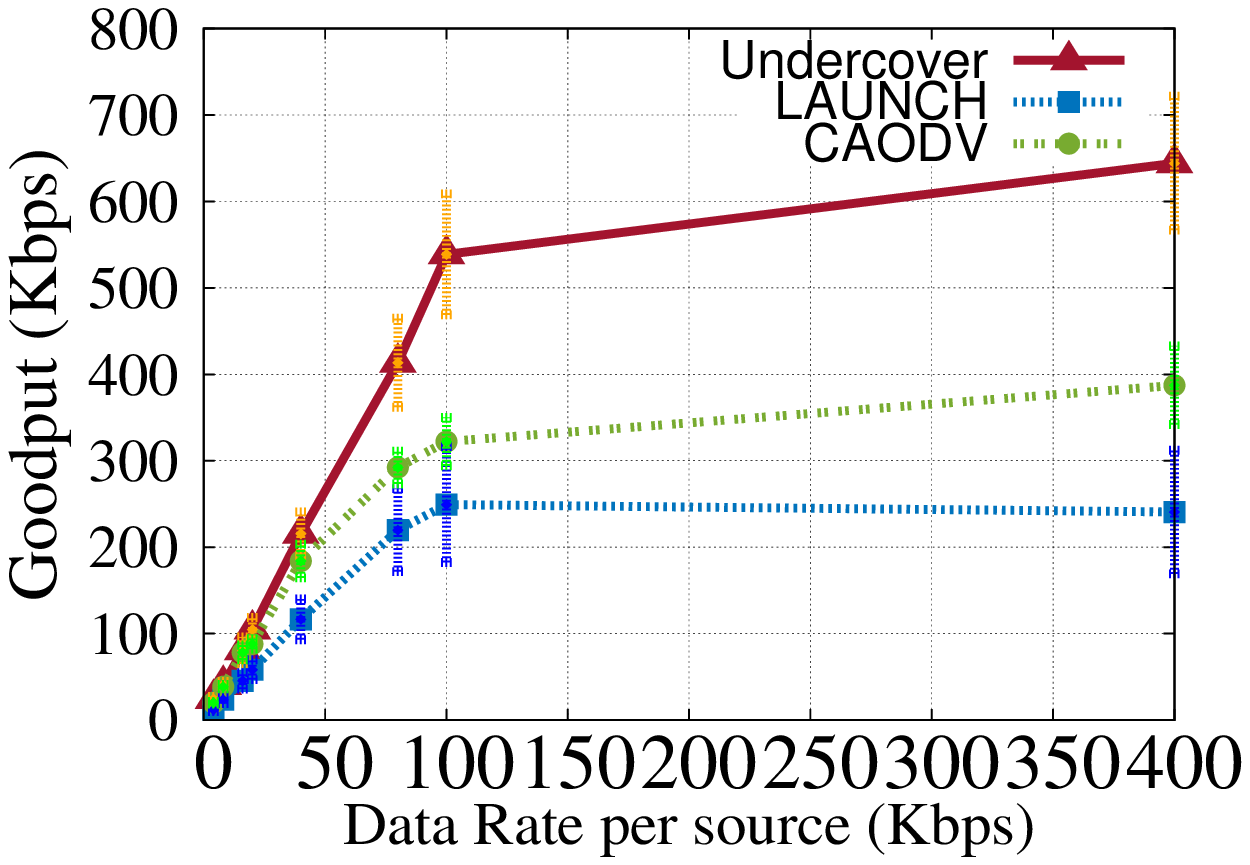}
	\caption{Goodput}
	\end{subfigure}
	\begin{subfigure}[t]{0.23\textwidth}
	\centering
    \includegraphics[width=1.7in]{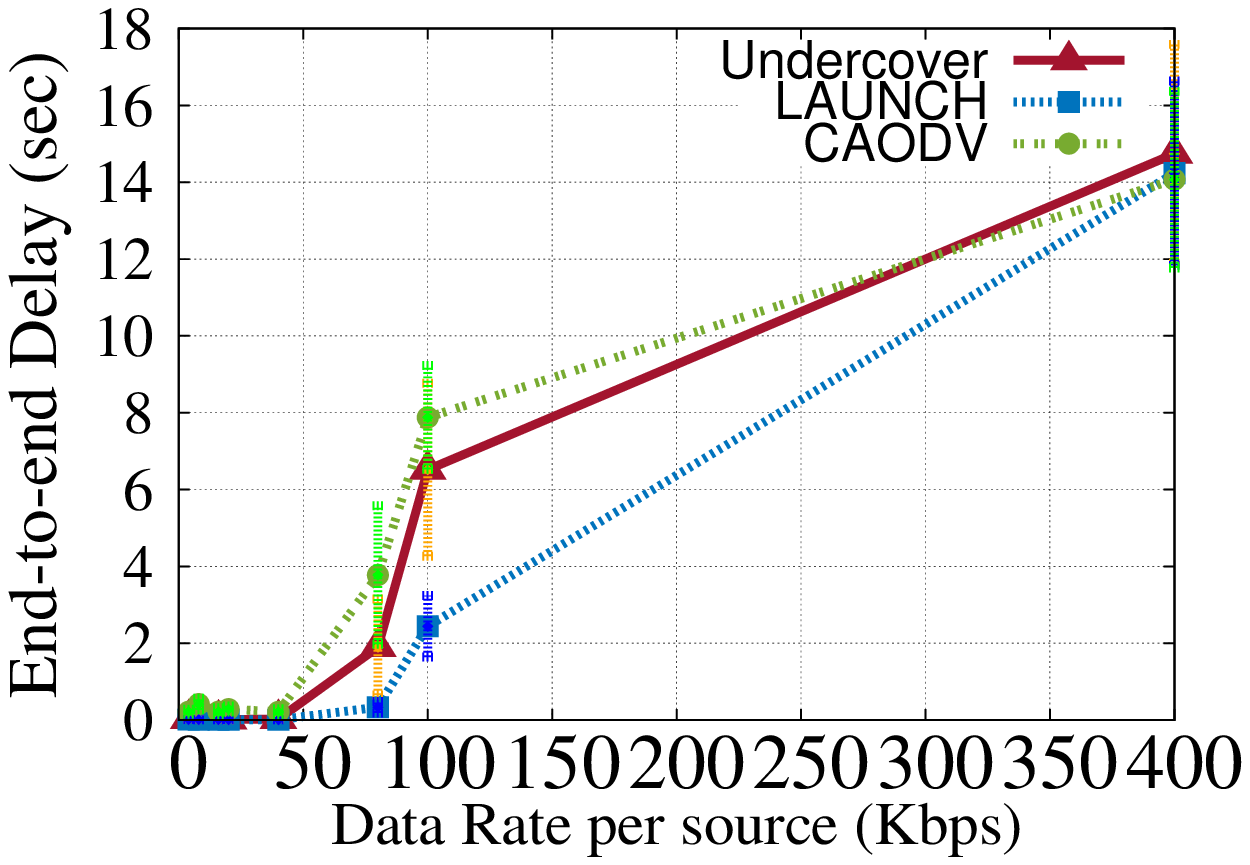}
	\caption{Average End-to-end Delay}
	\end{subfigure}
\caption{Effect of changing data rate per source on network performance.}
\label{fig::rate}
\end{figure}
 
\subsubsection{Changing Data Rate Per Source} 
Figures \ref{fig::rate} shows how \sys{} beats other routing protocols for different data rates. System performance increases in terms of the goodput but degrades in terms of the end-to-end delay for higher data rates. Some of the observations that are mentioned previously can be noted also in this figure.

\begin{figure}[!t]
\centering
	\begin{subfigure}[t]{0.23\textwidth}
	\centering
    \includegraphics[width=1.7in]{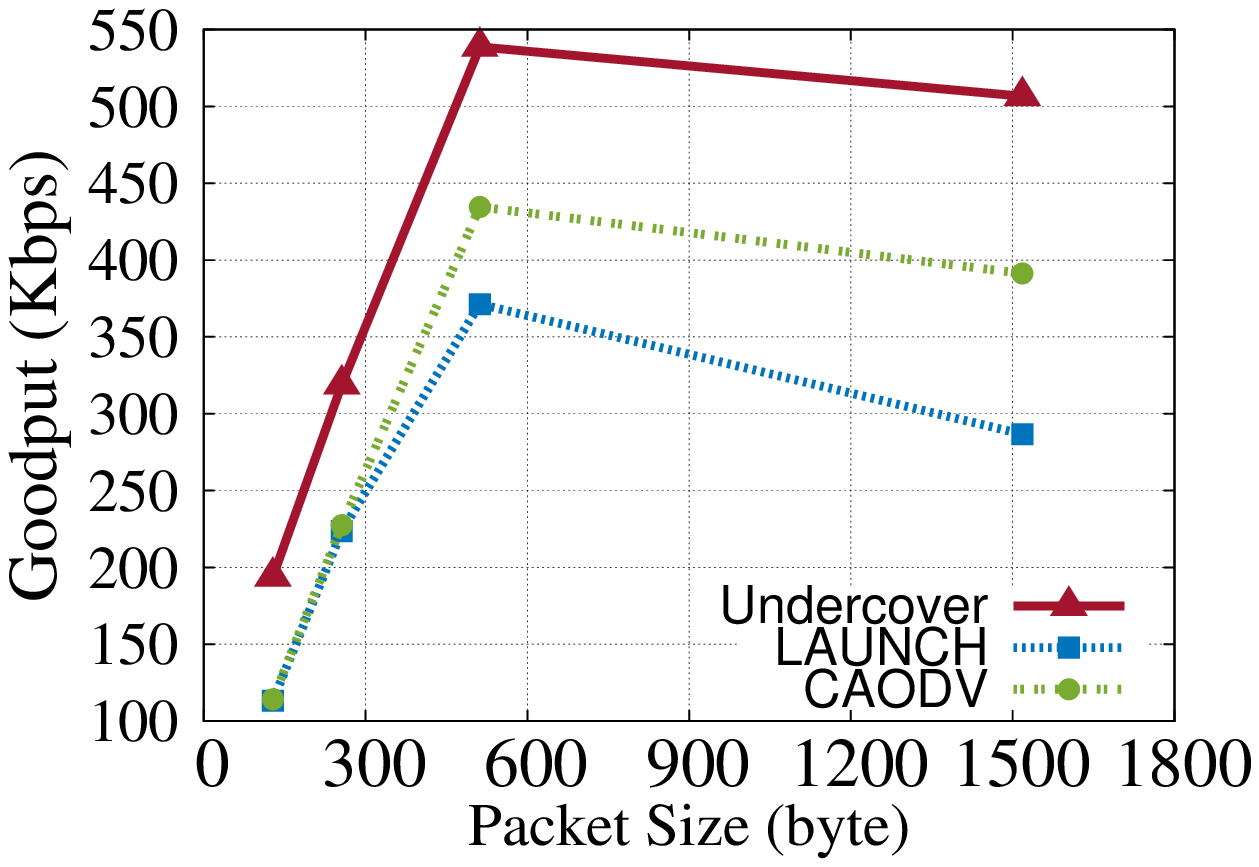}
	\caption{Goodput}
	\label{fig::pktsizethrough}
	\end{subfigure}
	\begin{subfigure}[t]{0.23\textwidth}
	\centering
    \includegraphics[width=1.7in]{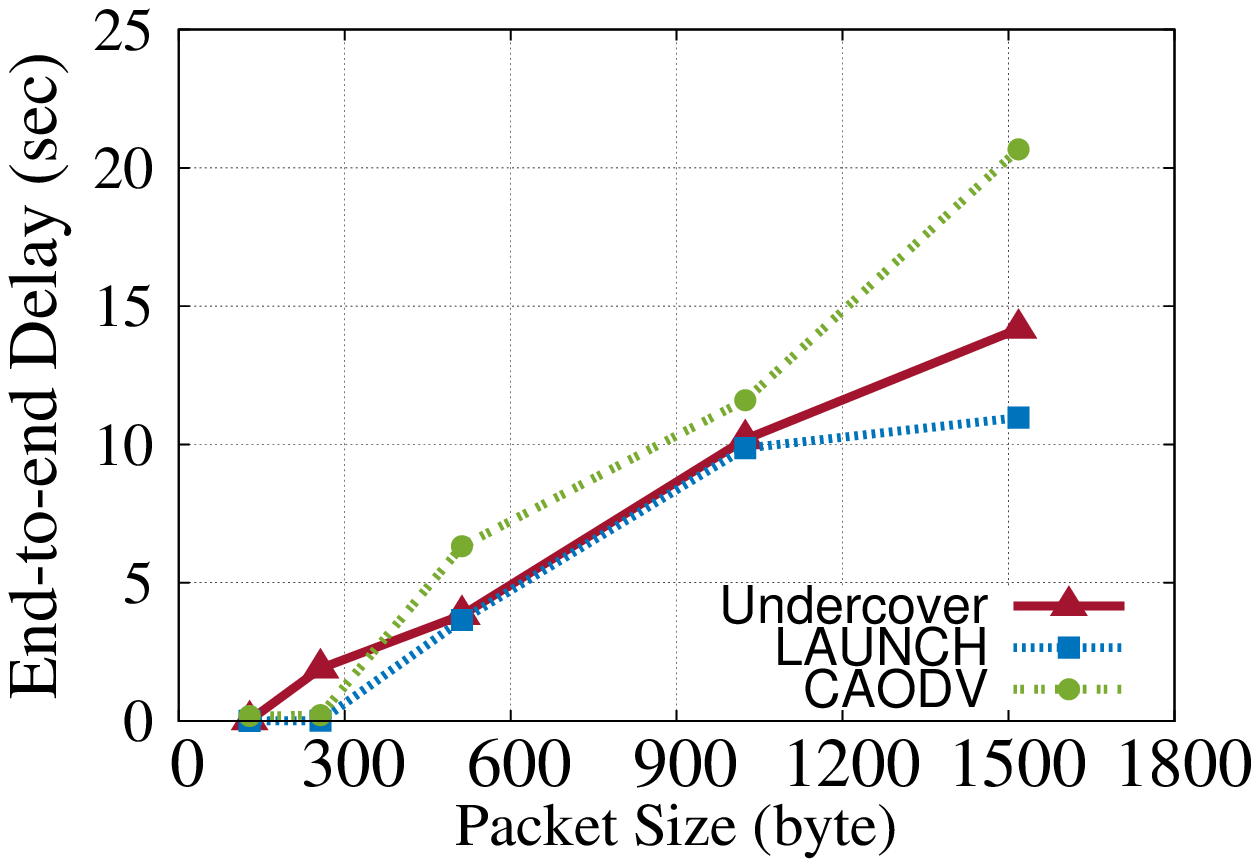}
	\caption{Average End-to-end Delay}
	\label{fig::pktsizedelay}
	\end{subfigure}
\caption{Effect of changing packet size on network performance.}
\label{fig::pktsize}
\end{figure}
 
\subsubsection{Changing Packet Size} 
Figures \ref{fig::pktsize} shows the effect of changing the packet size on the performance metrics for all used protocols for comparison. We can note generally that the average end-to-end delay (Figure \ref{fig::pktsizedelay}) increases as the packet size increases. This happens since any packet needs more time to reach its destination if its size increases. For Figure \ref{fig::pktsizethrough}, we can see that the goodput increases to some value for packet size then decreases again. The goodput increase at the first part of the figure happens due to delivering more data to final destinations when increasing the packet size. This case is similar to that of increasing data rate observed in Figure \ref{fig::rate}. However, the goodput decreases after that since some packets can not reach their destinations due to the introduced activity of PUs which preempt the sending process. So, these packets are lost and less data are then transmitted to their destinations safely. The last observation for this figure is that, we can see that \sys{} outperforms other protocols in terms of goodput when using any packet size.

 \begin{figure}
 \centering
  \includegraphics[width=0.4\textwidth]{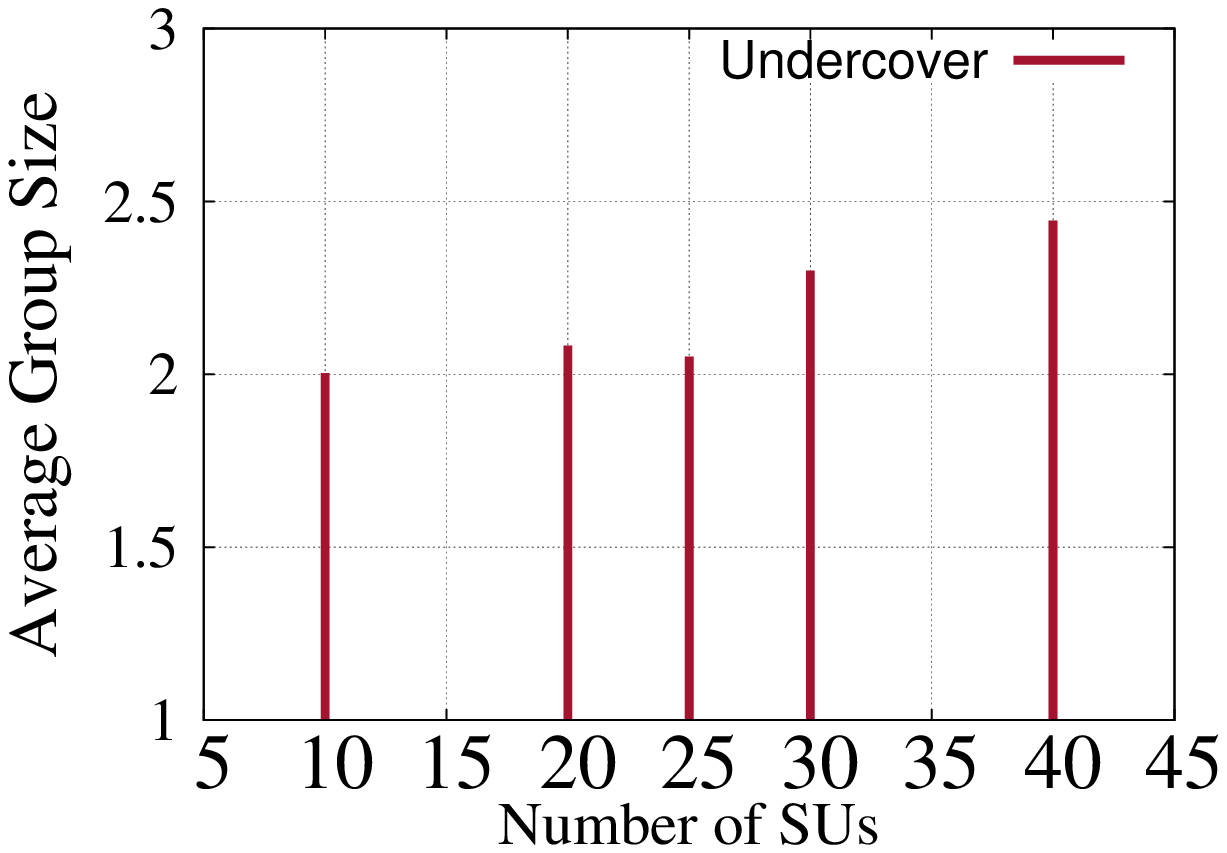}
  \caption{Average size of groups constructed by \sys{}.}
  \label{fig::gpsize}
 \end{figure}

\subsubsection{Average Goup Size}
Figure \ref{fig::gpsize} shows the average size for the groups constructed by \sys{}. It can be noted that as the number of SUs increases, the ability to construct larger group size increases. However, in almost all cases, \sys{} prefers to construct small sized groups to decrease the interference to the least possible effect especially when the number of PUs in the region of node's transmission is low.

\begin{figure}[!t]
\centering
	\begin{subfigure}[t]{0.23\textwidth}
	\centering
    \includegraphics[width=1.7in]{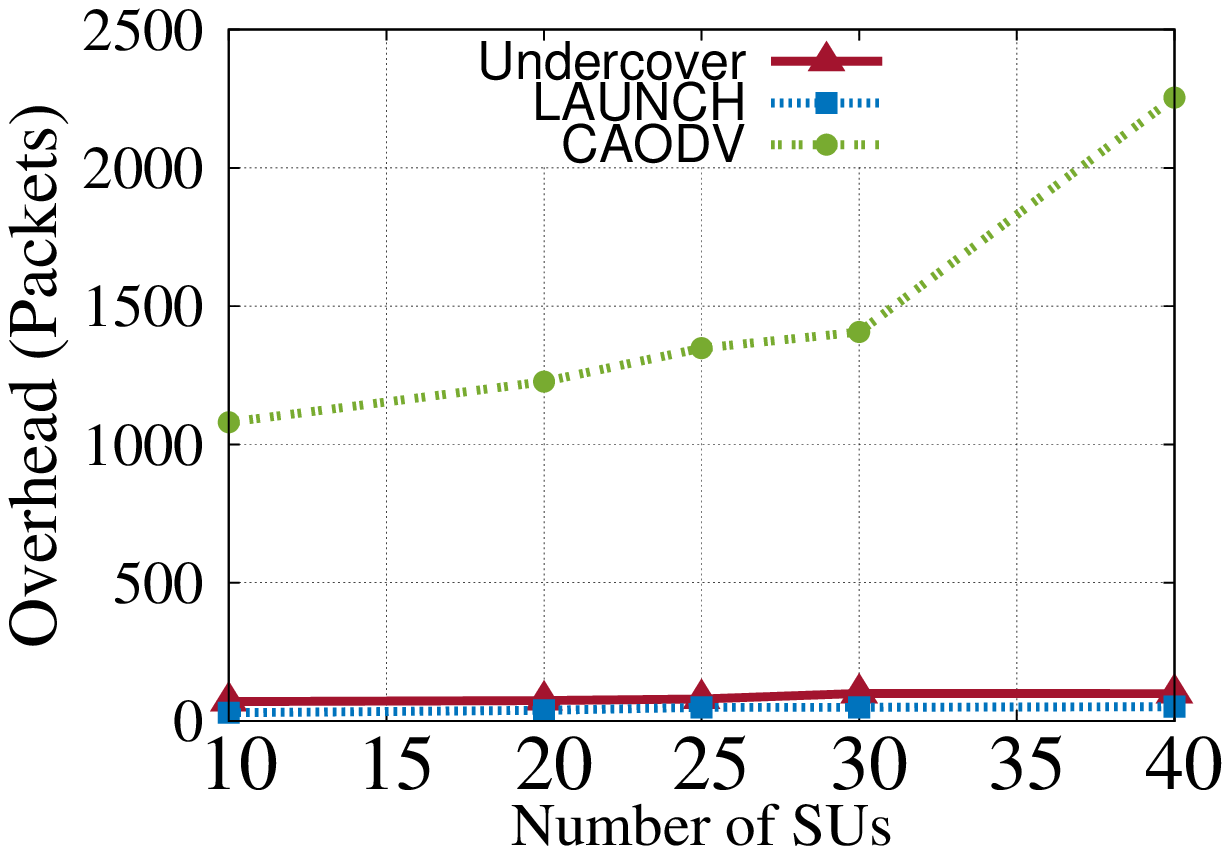}
	\caption{Comparison between the three routing protocols}
	\label{fig::overheadwocaodv}
	\end{subfigure}
	\begin{subfigure}[t]{0.23\textwidth}
	\centering
    \includegraphics[width=1.7in]{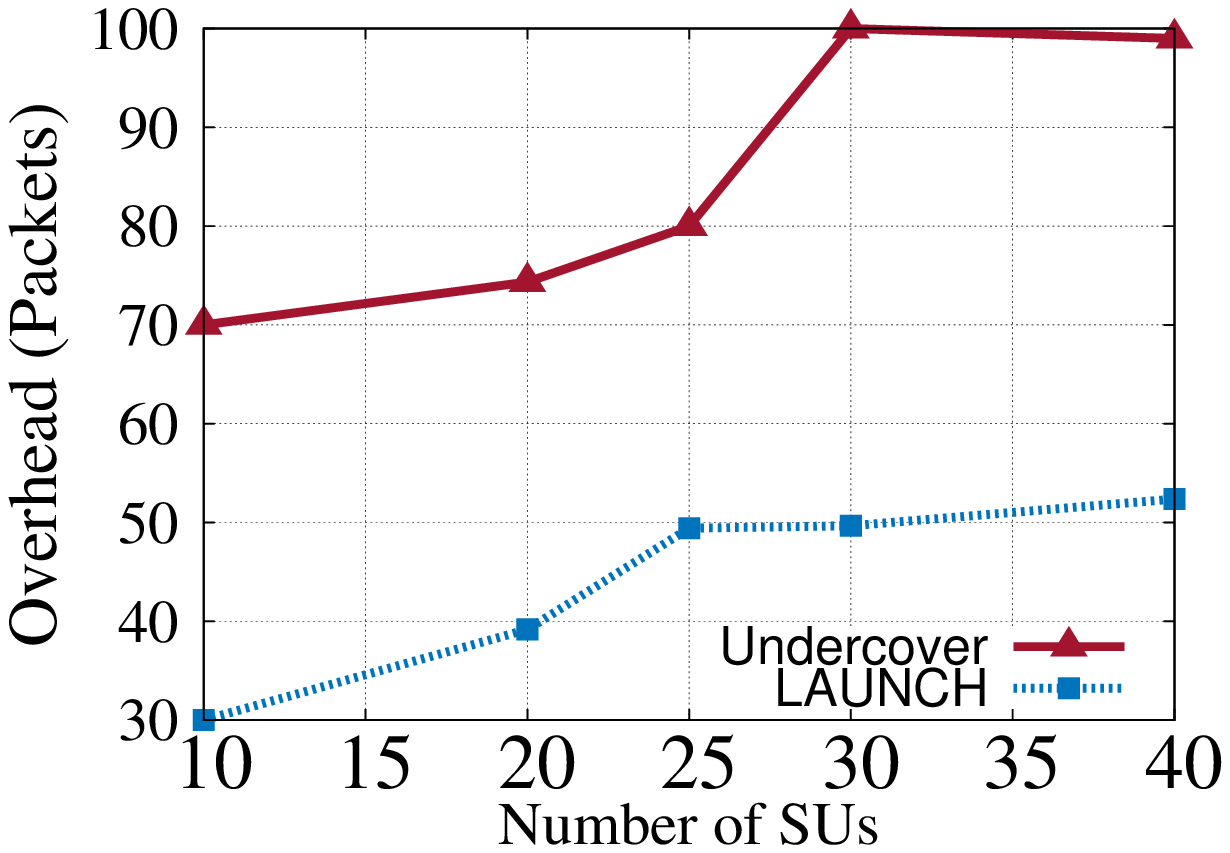}
	\caption{Zoomed figure on the part of low overhead protocols}
	\label{fig::overheadcaodv}
	\end{subfigure}
\caption{Effect of changing number of SUs on the routing overhead.}
\label{fig::overhead}
\end{figure}

\subsubsection{Routing Overhead}
Figure \ref{fig::overhead} shows the effect of increasing the number of SUs on the routing overhead. From Figure \ref{fig::overheadwocaodv}, we can see that CAODV has always higher overhead compared to both LAUNCH and \sys{}. This is due to the global routing approach used by CAODV which leads to having always a higher overhead compared to the local routing technique used by the other two protocols. Figure \ref{fig::overheadcaodv} compares both LAUNCH and \sys{} in terms of their added overhead packets. We can see that there is a nearly constant number of overhead packets added by \sys{} to help it in the groups construction process.

\section{Conclusion} \label{sec::conc}
In this paper, we proposed a new protocol \sys{} a cross
layering protocol that uses physical layer techniques in the routing layer.
Constructing cooperative groups and using beamforming can be used by secondary
users to send data normally even if primary users exist and are active through
nulling tranmissions at them. This property leads to better packet delivery
ratio for \sys{} than other protocols since they avoid the presence of primary users or
send when they are inactive. Thus, the ability to send simultaneously with primary users
opens a new degree of freedom that was not available before. Also, cooperative
groups idea is used to send signals in adhoc networks with better qualities. Thus, although our protocol is designed mainly for Cognitive Radio Networks, it proves to be useful also in adhoc networks. \sys{} is also designed to be interference-aware protocol so that it can take into
consideration the interference that constructed cooperative groups can do on
other routes and vice versa. NS2 is used for the evaluation in terms of the
achieved goodput and average end-to-end delay. \sys{} is compared against CAODV
which is a representative for the geographical protocols and LAUNCH as an
example from the location-aided routing protocols. \sys{} performs well achieving a goodput gain that reaches up to 250\% compared to other protocols. Also, it shows to have low overhead and reasonable end-to-end delay.

Future directions include finding a mathematical model for values in table \ref{table_threshold} and a way to improve the group construction time. Also, an important extension is to implement a dynamic protocol that assigns spectrum to the participating nodes in each group. We are also planning to test our protocol on some of the emerging testbeds like CRC \cite{crc}, CogFrame \cite{cogframe}, and CRESCENT \cite{crescent} to see the performance on real scenarios.

\bibliographystyle{IEEEbib}
\bibliography{myLib}

\begin{thebibliography}{10}

\bibitem{force2002report}
FCC Spectrum Policy~Task Force,
\newblock ``Report of the spectrum efficiency working group,'' 2002.

\bibitem{IoTRef}
``Facts and forecasts: Billions of things, trillions of dollars,'' 2014
  (accessed November, 2015).

\bibitem{crisis}
Peter Rysavy,
\newblock ``Spectrum crisis?,''
\newblock {\em Information Week Magazine}, pp. 23--30, 2009.

\bibitem{SDRMarket}
``Software-defined-radio market sees increased growth,'' 2014 (accessed
  November, 2015).

\bibitem{FCC-ruling}
Radio Magazine,
\newblock ``Fcc adopts rules for unlicensed use of television white spaces,''
\newblock {\em Retrieved at}, p.~4, 2008.

\bibitem{crn_metric_survey}
Moustafa Youssef, Mohammad Ibrahim, Mohamed Abdelatif, Lin Chen, and
  Athanasios~V Vasilakos,
\newblock ``Routing metrics of cognitive radio networks: A survey,''
\newblock {\em Communications Surveys \& Tutorials, IEEE}, vol. 16, no. 1, pp.
  92--109, 2014.

\bibitem{youssef2014routing}
Moustafa Youssef, Mohammad Ibrahim, Mohamed Abdelatif, Lin Chen, and
  Athanasios~V Vasilakos,
\newblock ``Routing metrics of cognitive radio networks: A survey,''
\newblock {\em Communications Surveys \& Tutorials, IEEE}, vol. 16, no. 1, pp.
  92--109, 2014.

\bibitem{pefkianakis2008samer}
Ioannis Pefkianakis, Starsky~HY Wong, and Songwu Lu,
\newblock ``Samer: spectrum aware mesh routing in cognitive radio networks,''
\newblock in {\em New Frontiers in Dynamic Spectrum Access Networks. DySPAN
  2008. 3rd Symposium on}. IEEE, pp. 1--5.

\bibitem{chowdhury2009search}
Kaushik~R Chowdhury and Marco~Di Felice,
\newblock ``Search: A routing protocol for mobile cognitive radio ad-hoc
  networks,''
\newblock {\em Computer Communications}, vol. 32, no. 18, pp. 1983--1997, 2009.

\bibitem{habak2013location}
Karim Habak, Mohammed Abdelatif, Hazem Hagrass, Karim Rizc, and Moustafa
  Youssef,
\newblock ``A location-aided routing protocol for cognitive radio networks,''
\newblock in {\em Computing, Networking and Communications (ICNC),
  International Conference on}. IEEE, 2013, pp. 729--733.

\bibitem{goldsmith2005wireless}
Andrea Goldsmith,
\newblock {\em Wireless communications},
\newblock Cambridge university press, 2005.

\bibitem{tse2005fundamentals}
David Tse and Pramod Viswanath,
\newblock {\em Fundamentals of wireless communication},
\newblock Cambridge university press, 2005.

\bibitem{van1988beamforming}
Barry~D Van~Veen and Kevin~M Buckley,
\newblock ``Beamforming: A versatile approach to spatial filtering,''
\newblock {\em IEEE assp magazine}, vol. 5, no. 2, pp. 4--24, 1988.

\bibitem{chen2014cooperative}
Xiaoming Chen, Hsiao-Hwa Chen, and Weixiao Meng,
\newblock ``Cooperative communications for cognitive radio networks—from
  theory to applications,''
\newblock {\em Communications Surveys \& Tutorials, IEEE}, vol. 16, no. 3, pp.
  1180--1192, 2014.

\bibitem{tao2012overview}
Xiaofeng Tao, Xiaodong Xu, and Qimei Cui,
\newblock ``An overview of cooperative communications,''
\newblock {\em IEEE Communications Magazine}, vol. 50, no. 6, pp. 65--71, 2012.

\bibitem{karmoose2013stability}
Mohammed Karmoose, Ahmed Sultan, and Moustafa Youssef,
\newblock ``Stability analysis in a cognitive radio system with cooperative
  beamforming,''
\newblock in {\em Wireless Communications and Networking Conference (WCNC)}.
  IEEE, 2013, pp. 637--642.

\bibitem{yi2012cooperative}
Tao Yi, Li~Guo, Kai Niu, Hongyan Cai, Jiaru Lin, and Wenbao Ai,
\newblock ``Cooperative beamforming in cognitive radio network with hybrid
  relay,''
\newblock in {\em 19th International Conference on Telecommunications (ICT)}.
  IEEE, 2012, pp. 1--5.

\bibitem{karmoose2013dead}
Mohammed Karmoose, Karim Habak, Mustafa ElNainay, and Moustafa Youssef,
\newblock ``Dead zone penetration protocol for cognitive radio networks,''
\newblock in {\em Wireless and Mobile Computing, Networking and Communications
  (WiMob), 9th International Conference on}. IEEE, 2013, pp. 529--536.

\bibitem{NS2}
S.~Floyd S.~McCanne,
\newblock ``{NS} network simulator,''
  http://www.w3schools.com/browsers/browsers\_os.asp.

\bibitem{lakshmanan2009diversity}
Sriram Lakshmanan and Raghupathy Sivakumar,
\newblock ``Diversity routing for multi-hop wireless networks with cooperative
  transmissions,''
\newblock in {\em IEEE SECON}, 2009, pp. 1--9.

\bibitem{collier1991transmission}
Christopher~J Collier and Robert~J Murray,
\newblock ``Transmission system for sending two signals simultaneously on the
  same communications channel,'' Dec.~17 1991,
\newblock US Patent 5,073,899.

\bibitem{khandani2005cooperative}
Amir~E Khandani, Eytan Modiano, Jinane Abounadi, and Lizhong Zheng,
\newblock ``Cooperative routing in wireless networks,''
\newblock in {\em Advances in Pervasive Computing and Networking}, pp. 97--117.
  Springer, 2005.

\bibitem{khandani2007cooperative}
Amir~Ehsan Khandani, Jinane Abounadi, Eytan Modiano, and Lizhong Zheng,
\newblock ``Cooperative routing in static wireless networks,''
\newblock {\em IEEE Transactions on Communications}, vol. 55, no. 11, pp.
  2185--2192, 2007.

\bibitem{li2006energy}
Fulu Li, Kui Wu, and Andrew Lippman,
\newblock ``Energy-efficient cooperative routing in multi-hop wireless ad hoc
  networks,''
\newblock in {\em 25th IEEE International Performance, Computing, and
  Communications Conference, IPCCC 2006.}, pp. 8--pp.

\bibitem{ibrahim2008distributed}
A~Ibrahim, Zhu Han, and KJ~Ray Liu,
\newblock ``Distributed energy-efficient cooperative routing in wireless
  networks,''
\newblock {\em IEEE Transactions on Wireless Communications}, vol. 7, no. 10,
  pp. 3930--3941, 2008.

\bibitem{jakllari2007cross}
Gentian Jakllari, Srikanth~V Krishnamurthy, Michalis Faloutsos, Prashant~V
  Krishnamurthy, and Ozgur Ercetin,
\newblock ``A cross-layer framework for exploiting virtual miso links in mobile
  ad hoc networks,''
\newblock {\em IEEE Transactions on Mobile Computing}, vol. 6, no. 6, pp.
  579--594, 2007.

\bibitem{lakshmanan2012proteus}
Sriram Lakshmanan and Raghupathy Sivakumar,
\newblock ``Proteus: Multiflow diversity routing for wireless networks with
  cooperative transmissions,''
\newblock {\em Mobile Computing, IEEE Transactions on}, vol. 12, no. 6, pp.
  1146--1159, 2013.

\bibitem{sharma2010cooperative}
Sushant Sharma, Yi~Shi, Y~Thomas Hou, Hanif~D Sherali, and Sastry Kompella,
\newblock ``Cooperative communications in multi-hop wireless networks: Joint
  flow routing and relay node assignment,''
\newblock in {\em INFOCOM, Proceedings}. IEEE, 2010, pp. 1--9.

\bibitem{zhang2009cooperative}
Qian Zhang, Juncheng Jia, and Jin Zhang,
\newblock ``Cooperative relay to improve diversity in cognitive radio
  networks,''
\newblock {\em IEEE Communications Magazine}, vol. 47, no. 2, pp. 111--117,
  2009.

\bibitem{ding2010distributed}
Lei Ding, Tommaso Melodia, Stella~N Batalama, and John~D Matyjas,
\newblock ``Distributed routing, relay selection, and spectrum allocation in
  cognitive and cooperative ad hoc networks,''
\newblock in {\em 2010 7th Annual IEEE Communications Society Conference on
  Sensor Mesh and Ad Hoc Communications and Networks (SECON)}, pp. 1--9.

\bibitem{sheu2012cooperative}
Jang-Ping Sheu and In-Long Lao,
\newblock ``Cooperative routing protocol in cognitive radio ad-hoc networks,''
\newblock in {\em IEEE Wireless Communications and Networking Conference
  (WCNC)}, 2012, pp. 2916--2921.

\bibitem{GPS99}
P.~Enge and P.~Misra,
\newblock ``{Special issue on GPS: The Global Positioning System},''
\newblock {\em Proceedings of the IEEE}, 1999.

\bibitem{GSM_loc4}
M.~Ibrahim and M.~Youssef,
\newblock ``{CellSense: An Accurate Energy-Efficient GSM Positioning System},''
\newblock {\em Vehicular Technology, IEEE Transactions on}, vol. 61, no. 1, pp.
  286--296, 2012.

\bibitem{punch}
Arsany {Guirguis}, Raymond {Guirguis}, and Moustafa {Youssef},
\newblock ``Primary user-aware network coding for multi-hop cognitive radio
  networks,''
\newblock in {\em Global Telecommunications Conference (GLOBECOM)}. IEEE, 2014.

\bibitem{de1997maximum}
Elisabeth De~Carvalho and Dirk~TM Slock,
\newblock ``Maximum-likelihood blind fir multi-channel estimation with gaussian
  prior for the symbols,''
\newblock in {\em IEEE International Conference on Acoustics, Speech, and
  Signal Processing ICASSP-97}, 1997, vol.~5, pp. 3593--3596.

\bibitem{khop}
Arsany {Guirguis}, Mohamed {Ibrahim}, Karim {Seddik}, Khaled {Harras}, Fadel
  {Digham}, and Moustafa {Youssef},
\newblock ``Primary user aware k-hop routing for cognitive radio networks,''
\newblock in {\em Global Telecommunications Conference (GLOBECOM)}. IEEE, 2015.

\bibitem{crextension}
``Cognitive radio extension.,''
\newblock {\em [Online]. Available: http://stuweb.ee.mtu.edu/ljialian/}.

\bibitem{gomadam2008approaching}
Krishna Gomadam, Viveck~R Cadambe, Syed Jafar, et~al.,
\newblock ``Approaching the capacity of wireless networks through distributed
  interference alignment,''
\newblock in {\em Global Telecommunications Conference.} IEEE, 2008, pp. 1--6.

\bibitem{yu2002models}
Kai Yu and Bj{\"o}rn Ottersten,
\newblock ``Models for mimo propagation channels: a review,''
\newblock {\em Wireless Communications and Mobile Computing}, vol. 2, no. 7,
  pp. 653--666, 2002.

\bibitem{caodv}
Angela~Sara Cacciapuoti, Cosimo Calcagno, Marcello Caleffi, and Luigi Paura,
\newblock ``Caodv: Routing in mobile ad-hoc cognitive radio networks,''
\newblock in {\em Wireless Days (WD), IFIP}. IEEE, 2010, pp. 1--5.

\bibitem{Perkins:2003:AHO:RFC3561}
C.~Perkins, E.~Belding-Royer, and S.~Das,
\newblock ``Ad hoc on-demand distance vector ({AODV}) routing,'' 2003.

\bibitem{crc}
``The cognetive radio cloud (crc) testbed,'' [Online]
  http://smartci.alexu.edu.eg/crc.

\bibitem{cogframe}
Ahmed Saeed, Mohammad Ibrahim, Khaled Harras, Moustafa Youssef, et~al.,
\newblock ``A low-cost large-scale framework for cognitive radio routing
  protocols testing,''
\newblock in {\em Communications (ICC), International Conference on}. IEEE,
  2013, pp. 2900--2904.

\bibitem{crescent}
Abdelrahman Asal, Ahmad Mamdouh, Ahmed Salama, Moamen Elgendy, Moamen Mokhtar,
  Muhammed Elsayed, and Moustafa Youssef,
\newblock ``Crescent: a modular cost-efficient open-access testbed for
  cognitive radio networks routing protocols,''
\newblock in {\em Proceedings of the 19th annual international conference on
  Mobile computing \& networking}. ACM, 2013, pp. 179--182.

\end{thebibliography}

\end{document}